\documentclass[10pt,onecolumn]{IEEEtran}       % onecolumn (second format)
\usepackage{amsfonts,amssymb,amsmath}
\usepackage{enumitem}
\usepackage{indentfirst,latexsym,amscd,algorithmic}
\usepackage{amsbsy,caption}
\usepackage[para]{threeparttable}
\usepackage{graphicx,multirow,bm}
\usepackage{caption}
\usepackage{color}
\usepackage{array}
\usepackage{subfigure}
\usepackage[ruled,vlined,commentsnumbered]{algorithm2e}
\usepackage{pifont,marvosym}
\usepackage{tikz}
\usepackage{longtable}
\usepackage{pdfpages}
\usepackage{geometry} 
\usepackage[justification=centering]{caption}
\usepackage[noadjust]{cite}

\usepackage{floatrow}
\newfloatcommand{capbtabbox}{table}[][\FBwidth]

\if@mathematic
   \def\vec#1{\ensuremath{\mathchoice
                     {\mbox{\boldmath$\displaystyle\bf{#1}$}}
                     {\mbox{\boldmath$\textstyle\bf{#1}$}}
                     {\mbox{\boldmath$\scriptstyle\bf{#1}$}}
                     {\mbox{\boldmath$\scriptscriptstyle\bf{#1}$}}}}
\else
   \def\vec#1{\ensuremath{\mathchoice
                     {\mbox{\boldmath$\displaystyle#1$}}
                     {\mbox{\boldmath$\textstyle#1$}}
                     {\mbox{\boldmath$\scriptstyle#1$}}
                     {\mbox{\boldmath$\scriptscriptstyle#1$}}}}
\fi

\allowdisplaybreaks[4]

\newcommand{\B}{{\cal B}}
\newcommand{\X}{{\cal X}}
\newcommand{\T}{{\cal T}}
\newcommand{\C}{{\cal C}}
\newcommand{\G}{{\cal G}}
\newcommand{\F}{{\cal F}}
\newcommand{\A}{{\cal A}}
\newcommand{\Hh}{{\cal H}}
\newcommand{\Pp}{{\cal P}}
\newcommand{\Y}{{\cal Y}}

\newtheorem{theorem}{Theorem}[section]
\newtheorem{definition}{Definition}[section]
\newtheorem{lemma}{Lemma}[section]
\newtheorem{corollary}[theorem]{Corollary}

\newtheorem{example}{Example}
\newtheorem{construction}{Construction}
\newtheorem{proposition}{Proposition}[section]

\definecolor{orange}{rgb}{1.0, 0.5, 0.0}
\definecolor{robin}{rgb}{0.0, 0.8, 0.8}

\begin{document}

\title{Signature codes for weighted binary adder channel and multimedia fingerprinting\\
\author{Jinping Fan, Yujie Gu, Masahiro Hachimori, and Ying Miao}
\thanks{J. Fan is with the Department of Policy and Planning Sciences, Graduate School of Systems and Information Engineering, University of Tsukuba, Tsukuba, Ibaraki 305-8573, Japan (e-mail: j.fan.math@gmail.com).}
\thanks{Y. Gu is with the Department of Electrical Engineering-Systems, Tel Aviv University, Tel Aviv, Israel and the Faculty of Information Science and Electrical Engineering, Kyushu University, Fukuoka, Japan (e-mail: guyujie2016@gmail.com).}
\thanks{M. Hachimori is with the Faculty of Engineering, Information and Systems, University of Tsukuba, Tsukuba, Ibaraki 305-8573, Japan (e-mail: hachi@sk.tsukuba.ac.jp).}
\thanks{Y. Miao is with the Faculty of Engineering, Information and Systems, University of Tsukuba, Tsukuba, Ibaraki 305-8573, Japan (e-mail: miao@sk.tsukuba.ac.jp). Research supported by JSPS Grant-in-Aid for Scientific Research (B) under Grant No. 18H01133.}
}

\date{}

\maketitle

\begin{abstract}
In this paper, we study {\color{black}binary signature codes} for the weighted binary adder channel (WbAC) and collusion-resistant multimedia fingerprinting. Let $A(n,t)$ denote the maximum size of a $t$-signature code of length $n$, and $A(n,w,t)$ denote the maximum size of a $t$-signature code of length $n$ and {\color{black}constant-weight} $w$. First, we derive asymptotic and general upper bounds {\color{black}on} $A(n,t)$ by relating signature codes to $B_t$ codes and bipartite graphs with large girth respectively, and also show the upper bounds are tight for certain cases. Second, we determine the exact values of $A(n,2,2)$ and $A(n,3,2)$ for infinitely many $n$ by connecting signature codes with $C_4$-free graphs and union-free families, respectively. Third, we provide two explicit constructions for $t$-signature codes which have efficient decoding algorithms and applications to two-level signature codes. Furthermore, we show from a geometric viewpoint that there does not exist any binary code with complete traceability for noisy WbAC and multimedia fingerprinting. A new type of signature codes with a weaker requirement than complete traceability is introduced for the noisy scenario.
\end{abstract}


\section{Introduction}
\label{sec:1}
The advancement of multimedia technologies with the development of communication networks has led to a tremendous use of multimedia {\color{black}content} such as images, videos and so on. However, such an advantage also poses the challenging task of resisting unauthorized redistribution of multimedia {\color{black}content}. Multimedia fingerprinting is a technique to protect continuous copyrighted data \cite{LTWWZ,TWWL} and several types of anti-collusion codes for multimedia fingerprinting have been investigated in recent decades, see \cite{B15,CFJLM,CM,EFKL2016,EFKL,GG,JCM} for example.

As in \cite{LTWWZ,TWWL}, suppose that the multimedia content is represented as a real-valued vector $\vec{x}=(x(1),x(2),\ldots,$ $x(m))\in\mathbb{R}^m$, called the \textit{host signal}. To prevent unauthorized redistribution of $\vec{x}$ outside of $M$ authorized users, the dealer constructs a set of watermarks, also called \textit{fingerprints}, using a linear modulation scheme based on $n\le m$ noise-like orthonormal signals $\F=\{\vec{f}_i\in\mathbb{R}^m:\ 1\le i\le n\}$. The set $\F$ is known to the dealer but unknown to {\color{black}any user}. The fingerprint $\vec{w}_j$ of the $j$-th authorized user, $1\le j\le M$, is represented as
\begin{equation}\label{eq-represent-w}
\vec{w}_j=\sum\limits_{i=1}^n c_j(i)\vec{f}_i,
\end{equation}
where $c_j(i)\in\{-1,1\}$ for \textit{antipodal modulation} and $c_j(i)\in\{0,1\}$ for \textit{on-off keying type of modulation} \cite{LTWWZ}. In this paper, we concentrate on the on-off keying type of modulation, that is, $c_j(i)\in\{0,1\}$. Notice that there exists a one-to-one correspondence between $\vec{w}_j$ and
$\vec{c}_j=(c_j(1),c_j(2),\ldots,c_j(n))\in\{0,1\}^n$
due to the linear independence of $\{\vec{f}_i: 1\le i\le n\}$. Hence, designing a collection of fingerprints with desired properties is equivalent to designing a set of binary {\color{black}vectors of length $n$}, $\{\vec{c}_j: 1\le j\le M\}\subseteq\{0,1\}^n$, with the corresponding properties. {\color{black}Moreover, if for any $1\le j\le M$, $\vec{c}_j$ has the same number of non-zero coordinates (i.e., constant-weight), then each user's fingerprint defined by (\ref{eq-represent-w}) is considered to have the same representation \textit{sparsity} under $\F=\{\vec{f}_1,\vec{f}_2,\ldots,\vec{f}_n\}$.}

In the above setting, the dealer distributes to the $j$-th authorized user the signal
\begin{equation*}
  \vec{y}_j=\vec{x}+\vec{w}_j
\end{equation*}
under the assumption that {\color{black}$\|\vec{x}\|\gg\|\vec{w}_j\|$,\footnote{$\|\cdot\|$ is the Euclidean norm.}} which ensures that the embedded {\color{black}fingerprint does} not make a significant change to the host signal. A group of malicious users (\textit{coalition}) may come together and create a forged copy from their fingerprinted contents, but under the \textit{Multimedia Marking Assumption} \cite{EFKL}, they cannot manipulate the orthonormal signals $\vec{f}_i$, $1\le i\le n$. In a \textit{linear attack}, a coalition $J\subseteq\{1,2,\ldots,M\}$ creates a forged copy $\hat{\vec{y}}$ by taking a linear combination of their copies $\vec{y}_j$ with some real-valued coefficients $\lambda_j$, that is,
\begin{equation}\label{eq-linearattack}
  \hat{\vec{y}}=\sum\limits_{j\in J}\lambda_j\vec{y}_j{\color{black}=\sum_{j\in J}\lambda_j(\vec{x}+\vec{w}_j),}
\end{equation}
where $\sum_{j\in J}\lambda_j=1$, and $\lambda_j>0$ for all $j\in J$ since we consider that all users in the coalition (\textit{colluders}) make contributions to the forged copy $\hat{\vec{y}}$. Equivalently,
\begin{equation*}
  \hat{\vec{y}}=\vec{x}+\sum\limits_{j\in J}\lambda_j\vec{w}_j,
\end{equation*}
where the coefficients $\lambda_j$ are chosen by the coalition $J$ but unknown to the dealer. When $|J|=t$ and $\lambda_j=1/t$ for all $j\in J$, the linear attack is known as the \textit{averaging attack}. Once a forged copy $\hat{\vec{y}}$ is confiscated, the dealer can calculate
\begin{equation}\label{$r_k$}
  r(k)=\langle\hat{\vec{y}}-\vec{x},\vec{f}_k\rangle=\Big\langle\sum\limits_{i=1}^n\sum\limits_{j\in J}\lambda_jc_j(i)\vec{f}_i,\vec{f}_k\Big\rangle=\sum\limits_{j\in J}\lambda_jc_j(k)
\end{equation}
for $1\le k\le n$ where $\langle\cdot,\cdot\rangle$ denotes the inner product, and obtain
\begin{equation}\label{$r$}
  \vec{r}=(r(1),\ldots,r(n))=\sum_{j\in J}\lambda_j\vec{c}_j.
\end{equation}
{\color{black}Obviously, $\vec{r}\in\mathbb{R}^n$ and $0\le r(k)\le 1$ for any $1\le k\le n$.} To resist the linear attack, the dealer aims to identify the coalition $J$ based on the result $\vec{r}$ {\color{black}even without any knowledge of $\lambda_j$, $j\in J$,} which requires that the set $\{\vec{c}_j: 1\le j\le M\}\subseteq\{0,1\}^n$ has some desired properties.

{\color{black}
In the previous work, Cheng and Miao \cite{CM} considered a \textit{discretized model} of multimedia fingerprinting. In their setting, the \textit{discretization} of the result $\vec{r}$ was considered, which can be represented by a mapping $\sigma$ on $\vec{r}$: $\sigma(\vec{r})=\sigma(r(1))\times\cdots\times\sigma(r(n))$ where
\begin{equation*}
  \sigma(r(k))=\left\{\begin{array}{ll}
  \{r(k)\}, & r(k)=0\ \text{or}\ 1,  \\
  \{0,1\}, & 0<r(k)<1,
\end{array}\right.
\end{equation*}
for any $1\le k\le n$. Since $\sum_{j\in J}\lambda_j=1$ and $\lambda_j>0$ for all $j\in J$, by (\ref{$r_k$}), it is easy to see that $\sigma(r(k))=\{c_j(k): j\in J\}$ for any $1\le k\le n$. Under this discretized model of multimedia fingerprinting, several types of codes have been proposed. In~\cite{CM}, Cheng and Miao introduced \textit{separable codes} (SCs) to identify the coalition set $J$ based on $\sigma(\vec{r})$, which were further investigated in \cite{B15,EFKL2016,GG} and more. It was proved in~\cite{CM} that \textit{frameproof codes} (FPCs), introduced by Boneh and Shaw in~\cite{BS}, have the ability of identifying the whole coalition set $J$ and the tracing algorithm based on an FPC is more efficient than that based on an SC. However the requirements of an FPC {\color{black}are stronger than those} of an SC. To combine the advantages of FPCs and SCs, Jiang et al. \cite{JCM} introduced \textit{strongly separable codes} (SSCs) which lie between FPCs and SCs, and the tracing algorithm based on an SSC is as efficient as that based on an FPC. Also under the discretized model, Cheng et al.~\cite{CFJLM} considered the problem of identifying at least one colluder and introduced codes with the \textit{identifiable parent property} for multimedia fingerprinting, whose largest code size and efficient decoding algorithm were later investigated in~\cite{GCKM,JGC} and more.

As can be seen, the above discretized model does not explore the randomness of $\lambda_j$ in the attack model~\eqref{eq-linearattack}.
In this paper,} we would like to go back to the original problem of designing $\{\vec{c}_j: 1\le j\le M\}\subseteq\{0,1\}^n$ to trace back to the whole coalition $J$ based on the result $\vec{r}$ even though $\lambda_j$, $j\in J$ are unknown, that is, designing an \textit{anti-collusion code} with \textit{complete traceability} in multimedia fingerprinting, where the complete traceability refers to the ability of tracing back to all colluders, i.e., the whole coalition $J$.

In \cite{EFKL}, Egorova et al. showed that this problem is in fact equivalent to the problem of designing a \textit{$t$-signature code} {\color{black}for} the \textit{weighted binary adder channel} (WbAC). Suppose that $M$ users would like to communicate with the same destination through a shared WbAC in the \textit{multiple-access communication system}, among which at most $t$ users are \textit{active} simultaneously. To communicate successfully, each user $j$ is encoded to a unique vector $\vec{c}_j\in\{0,1\}^n$, $1\le j\le M$. If the users in $J\subseteq[M]$ are active at the same time, they input their vectors simultaneously into the WbAC. The output in the destination is a vector $\vec{r}$ as in (\ref{$r$}), where $\lambda_j$ plays the role of weight depending only on the channel but unknown to all encoders and the decoder. To identify all the active users in set $J$ using the corresponding channel output $\vec{r}$, it is required to design $\{\vec{c}_j: 1\le j\le M\}\subseteq\{0,1\}^n$ with some desired properties. The case that $\lambda_j>0$, $j\in J$ are real numbers in WbAC was first investigated by Mathys \cite{M}. Egorova et al. \cite{EFKL} first considered the scenario that $\lambda_j>0$, $j\in J$ are real numbers such that $\sum_{j\in J}\lambda_j=1$. Under this setting, they observed that the WbAC is essentially a modification of the multimedia fingerprinting channel, {\color{black}and the problem of designing $\{\vec{c}_j: 1\le j\le M\}\subseteq\{0,1\}^n$ to identify the set of active users from the output $\vec{r}$ of WbAC is equivalent to the problem of designing $\{\vec{c}_j: 1\le j\le M\}\subseteq\{0,1\}^n$ with complete traceability in multimedia fingerprinting where $\vec{r}$ is defined in (\ref{$r$}). Based on this observation, they introduced the concept of {\color{black}signature codes} for WbAC and multimedia fingerprinting and gave a direct construction for signature codes which yields a lower bound on {\color{black}the maximum size of} $t$-signature codes. As far as we know, there are no other results on signature codes except that shown in \cite{EFKL}.}

{\color{black}
In this paper, we would like to investigate signature codes from a combinatorial perspective. Specifically, we derive upper bounds on {\color{black}the maximum size of} signature codes by combinatorial methods, which improve the known results in \cite{EFKL} for certain cases. We also investigate the combinatorial properties of signature codes with constant-weight and establish their optimal constructions. Moreover, we provide two constructions for signature codes which are equipped with efficient decoding algorithms. In addition, we consider some variants of signature codes for the noisy scenario as well. Due to the equivalence between WbAC and multimedia fingerprinting as described in the last paragraph, in the following contents, we will discuss codes mainly with respect to multimedia fingerprinting for simplicity.
}

The remainder of this paper is organized as follows. {\color{black}In Section \ref{sec:signature code}, we review the definition of signature codes and give an equivalent description of signature codes from a geometric viewpoint. Known results on signature codes and relationships between signature codes and $B_t$ codes are also presented}. In Section \ref{sec:2}, we first relate $t$-signature codes to bipartite graphs without cycles of length less than or equal to $2t$ and then obtain general upper bounds for $t$-signature codes of length $n$ from the known results on bipartite graphs. Asymptotic upper bounds for $t$-signature codes are obtained based on the known results on $B_t$ codes. In Section \ref{sec:3}, we first show that a $2$-signature code with constant-weight $2$ is equivalent to a $C_4$-free graph, yielding exact  values of $A(n,2,2)$ and optimal constructions for $2$-signature codes with constant-weight $2$ for infinite values of $n$. Exact values of $A(n,3,2)$ are also derived and optimal constructions for $2$-signature codes with constant-weight $3$ are obtained for all $n$ with a few exceptions by investigating their combinatorial properties. In Section \ref{sec:4}, we provide two explicit constructions for $t$-signature codes and also investigate their decoding algorithms and applications. In Section \ref{sec:5}, we show from a geometric viewpoint that there does not exist any binary code with complete traceability for noisy multimedia fingerprinting, and introduce a new type of signature codes with a weaker requirement which could be used in noisy multimedia fingerprinting. Conclusion is drawn in Section \ref{sec:6}.

%{\color{black}
\section{\color{black}Preliminaries}
\label{sec:signature code}

In this section, we first present the definition of signature code introduced in \cite{EFKL}, and then give an equivalent description of signature codes from a geometric viewpoint. Known results on signature codes and relationships between signature codes and $B_t$ codes are also presented.

\subsection{Signature code}
\label{sec1,sub1:signature code}

Let $n,M$ and $q$ be positive integers, and $Q=\{0,1,\ldots,q-1\}$ be the alphabet. A set $\C=\{\vec{c}_1,\vec{c}_2,\ldots,\vec{c}_M\}\subseteq Q^n$ is called a \textit{$q$-ary code} of length $n$ and size $M$, or $(n,M,q)$ code. Each $\vec{c}_j=(c_j(1),c_j(2),\ldots,c_j(n))$ of $\C$ is called a \textit{codeword} and $c_j(i)$ is called the $i$-th \textit{coordinate} of $\vec{c}_j$. Let $\C$ be an $(n,M,q)$ code. Define $R=(\log_q M)/n$ as the \textit{code rate} of $\C$.
{\color{black}
The \textit{matrix representation} of $\C$ is an $n\times M$ matrix where each column vector is a codeword of $\C$. Denote $[n]=\{1,2,\ldots,n\}$. For any $I\subseteq[n]$ and $\vec{c}=(c(1),c(2),\ldots,c(n))\in\C$, we denote $\vec{c}|_I$ as the \textit{punctured codeword} of $\vec{c}$ restricted to $I$ by deleting all the coordinates $c(k)$ of $\vec{c}$ with $k\not\in I$, and denote $\C|_I=\{\vec{c}|_I: \vec{c}\in\C\}$ as the \textit{punctured code} of $\C$ restricted to $I$.
}

An $(n,M,2)$ code is also called a \textit{binary} code. Let $\C$ be a binary code. For any codeword $\vec{c}\in\C$, let $\mathrm{supp}(\vec{c})=\{i\in [n]: c(i)=1\}$ be the \textit{support} of $\vec{c}$. Denote $2^{[n]}$ as the family of all subsets of $[n]$, and $\binom{[n]}{w}$ as the family of all $w$-subsets of $[n]$. $\C$ has {\color{black}\textit{constant-weight}} $w$ if $|\mathrm{supp}(\vec{c})|=w$ for any $\vec{c}\in\C$.

The definition of signature code was stated in \cite{EFKL}.

{\color{black}
\begin{definition}\label{sig. code}{\rm(\cite{EFKL})}
{\color{black}
Let $\C$ be an $(n,M,2)$ code and $t\ge2$ be an integer. $\C$ is an $(n,M,2)$ \textit{$t$-signature code} if

for any two distinct subsets $\C_1,\C_2\subseteq\C$ with $1\le|\C_1|,|\C_2|\le t$, we have
\begin{equation*}
  \sum_{\vec{c}_j\in\C_1}\lambda_j\vec{c}_j\neq\sum_{\vec{c}_k\in\C_2}\lambda_k'\vec{c}_k
\end{equation*}
for any real numbers $\lambda_j,\lambda_k'>0$ such that $\sum_{\vec{c}_j\in\C_1}\lambda_j=\sum_{\vec{c}_k\in\C_2}\lambda_k'=1$.
}
\end{definition}
}

%\begin{remark}\label{remark-def}
%An equivalent description of Definition \ref{sig. code} is as follows. An $(n,M,2)$ code $\C$ is a $t$-signature code if for any two distinct subsets $\C_1,\C_2\subseteq\C$ with $1\le|\C_1|,|\C_2|\le t$, we have
%\begin{equation*}
%  \sum_{\vec{c}_j\in\C_1}\lambda_j\vec{c}_j\neq\sum_{\vec{c}_k\in\C_2}\lambda_k'\vec{c}_k
%\end{equation*}
%for any real numbers $\lambda_j,\lambda_k'>0$ such that $\sum_{\vec{c}_j\in\C_1}\lambda_j=\sum_{\vec{c}_k\in\C_2}\lambda_k'=1$.
%\end{remark}

We first generalize the definition of signature code by considering multi-subsets of $\C$ in the condition of Definition \ref{sig. code}. A \textit{multi-set} is a collection of elements which allows for multiple occurrences of the elements. The number of occurrence of each element in a multi-set is called the \textit{multiplicity} of this element. The \textit{size} of a multi-set is the sum of the multiplicities of all its elements. For any multi-set $S$, we call the set formed by all the elements in $S$ as the \textit{base set} of $S$. Two multi-sets are \textit{distinct} if their base sets are distinct, or there exists an element such that the multiplicity of this element in two multi-sets are different. A \textit{multi-subset} $S_1$ of a set $S_2$ is a multi-set such that the base set of $S_1$ is a subset of $S_2$.

\begin{example}
Let $S=\{1,2,3,4,5\}$ be a set. Then $S_1=\{1,1,2\}$ and $S_2=\{1,2,2\}$ are two distinct multi-subsets of $S$ with the same base set $\{1,2\}$. Let $S_3=\{1,2,2,3\}$. Then the base set of $S_3$ is $\{1,2,3\}$ which is distinct to that of $S_1$ and $S_2$.
\end{example}

We have the following equivalent but more general description of Definition \ref{sig. code}.

{\color{black}
\begin{proposition}\label{def-general-equiv}
An $(n,M,2)$ code $\C$ is a $t$-signature code if and only if for any two multi-subsets $\widehat{\C}_1=\{\vec{c}_{j_1},\ldots,\vec{c}_{j_s}\}$ and $\widehat{\C}_2=\{\vec{c}_{k_1},\ldots,\vec{c}_{k_m}\}$ of $\C$ with $s,m\le t$ and distinct base sets, we have
\begin{equation*}
  \sum_{h=1}^s\lambda_{j_h}\vec{c}_{j_h}\neq\sum_{l=1}^m\lambda_{k_l}'\vec{c}_{k_l}
\end{equation*}
for any real numbers $\lambda_{j_h},\lambda_{k_l}'>0$ such that $\sum_{h=1}^s\lambda_{j_h}=\sum_{l=1}^m\lambda_{k_l}'=1$.
\end{proposition}
\begin{IEEEproof}
{\color{black}
The sufficiency could be easily derived by Definition \ref{sig. code}. To show the necessity, we assume that $\C=\{\vec{c}_1,\cdots,\vec{c}_M\}$ is a $t$-signature code, but there exist two multi-subsets $\widehat{\C}_1=\{\vec{c}_{j_1},\ldots,\vec{c}_{j_s}\},\ \widehat{\C}_2=\{\vec{c}_{k_1},\ldots,\vec{c}_{k_m}\}$ of $\C$ with $s,m\le t$ and distinct base sets, and also exist $\lambda_{j_h},\lambda_{k_l}'>0$ with $\sum_{h=1}^s\lambda_{j_h}=\sum_{l=1}^m\lambda_{k_l}'=1$ such that
\begin{equation}\label{Eq:general sig.code-1}
  \sum_{h=1}^s\lambda_{j_h}\vec{c}_{j_h}=\sum_{l=1}^m\lambda_{k_l}'\vec{c}_{k_l}.
\end{equation}

By combining like terms in the left-hand and right-hand sides of (\ref{Eq:general sig.code-1}) respectively, we can obtain
\begin{equation*}
  \sum_{\vec{c}_j\in\C_1}\lambda_j\vec{c}_j=\sum_{\vec{c}_k\in\C_2}\lambda_k'\vec{c}_k
\end{equation*}
where $\C_1,\C_2$ are base sets of $\widehat{\C}_1,\widehat{\C}_2$ respectively satisfying $|\C_1|,|\C_2|\le t$ and $\C_1\neq\C_2$,
and $\lambda_j, \lambda'_k>0$ are real numbers satisfying $\sum_{\vec{c}_j\in\C_1}\lambda_j=\sum_{\vec{c}_k\in\C_2}\lambda_k'=1$. This is a contradiction to the assumption that $\C$ is a $t$-signature code (see Definition~\ref{sig. code}). The conclusion follows.
}
\end{IEEEproof}

%%%%%%%%%%%%%%%%%%%%%%%%%%%%%%%%%%%%%%%%%%%%%
%{\color{brown}(Is it the case that $\C_1,\C_2$ are base sets of $\widehat{\C}_1,\widehat{\C}_2$ respectively???)}

% Yes, you are correct.
%%%%%%%%%%%%%%%%%%%%%%%%%%%%%%%%%%%%%%%%%%%%%

Proposition \ref{def-general-equiv} shows that in Definition \ref{sig. code}, the condition ``two distinct subsets $\C_1,\C_2\subseteq\C$ with $1\le|\C_1|,|\C_2|\le t$'' could be equivalently replaced by ``two multi-subsets $\C_1,\C_2\subseteq\C$ of size no more than $t$ with distinct bases sets''. We remark that in the {\color{black}remainder} of this paper, $\C_1,\C_2\subseteq\C$ {\color{black}always refer to} subsets {\color{black}unless otherwise} specified.

}

{\color{black}
\subsection{A geometric perspective}
}

Signature codes can be interpreted geometrically. Let $\{0,1\}^n$ be the vertex-set of an $n$-dimensional hypercube. Then a binary code of length $n$ corresponds to a subset of vertices of an $n$-dimensional hypercube. For any subset of vertices $\C=\{\vec{c}_1,\vec{c}_2,\ldots,\vec{c}_m\}\subseteq\{0,1\}^n$, define the \textit{open convex polytope} of $\C$ as
\begin{equation}\label{open convex polytope}
\mathcal{P}(\C)=\Big\{\sum\limits_{i=1}^m \lambda_i\vec{c}_i:\ \sum\limits_{i=1}^m\lambda_i=1\ \text{and}\ \lambda_i> 0,\ \forall 1\le i\le m\Big\}.
\end{equation}
For any distinct $\C_1,\C_2\subseteq\{0,1\}^n$, the open convex polytopes of $\C_1$ and $\C_2$ do not \textit{cross} if
\begin{equation*}
  \mathcal{P}(\C_1)\cap \mathcal{P}(\C_2)=\emptyset.
\end{equation*}
Then we have

{\color{black}
\begin{proposition}\label{noiseless geometric}
{\color{black}An $(n,M,2)$ code $\C$ is a $t$-signature code if and only if for any distinct subsets $\C_1,\C_2\subseteq\C$ with $1\le|\C_1|,|\C_2|\le t$, we have $\mathcal{P}(\C_1)\cap \mathcal{P}(\C_2)=\emptyset$.}
\end{proposition}
}

Proposition \ref{noiseless geometric} can be immediately derived from Definition \ref{sig. code}. The following example illustrates our argument. %of P

\begin{example}\label{3-dim cube}
Consider the vertices of the $3$-dimensional hypercube as shown in Figure \ref{fig.1}. The open convex polytope formed by any two distinct vertices $v_i,v_j$ is the line segment between $v_i$ and $v_j$ except $v_i,v_j$ themselves. Let $\Hh_0=\{v_1,v_2,v_3,v_6,v_8\}$. It is easy to check that the open convex polytopes formed by any two vertices in $\Hh_0$ do not cross.
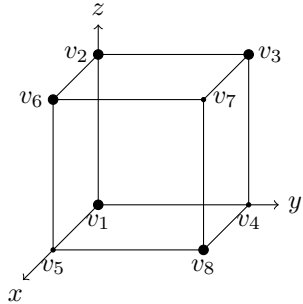
\begin{figure}
%\begin{center}
\begin{tikzpicture}[scale=2]
       \coordinate (A5) at (0, 0);
       \coordinate (A6) at (0, 1);
       \coordinate (A7) at (1, 1);
       \coordinate (A8) at (1, 0);
       \coordinate (A1) at (0.3, 0.3);
       \coordinate (A2) at (0.3, 1.3);
       \coordinate (A3) at (1.3, 1.3);
       \coordinate (A4) at (1.3, 0.3);

       \draw[] (A5) -- (A6);
       \draw[] (A6) -- (A7);
       \draw[] (A7) -- (A8);
       \draw[] (A8) -- (A5);
       %\draw[dashed] (A1) -- (B1);
       %\draw[dashed] (B1) -- (B2);
       \draw[] (A6) -- (A2);
       \draw[] (A2) -- (A3);
       \draw[] (A7) -- (A3);
       \draw[] (A8) -- (A4);
       \draw[] (A4) -- (A3);
       %\draw[dashed] (B1) -- (B4);

       \draw [->] (0.3,0.3) -- (1.5,0.3) ;
       \draw [->] (0.3,0.3) -- (0.3,1.5) ;
       \draw [->] (0.3,0.3) -- (-0.2,-0.2) ;

       \filldraw[black] (0,0) circle (0.4pt) node[anchor=west] {};
       \filldraw[black] (0,1) circle (0.9pt) node[anchor=west] {};
       \filldraw[black] (1,1) circle (0.4pt) node[anchor=west] {};
       \filldraw[black] (1,0) circle (0.9pt) node[anchor=west] {};
       \filldraw[black] (0.3,0.3) circle (0.9pt) node[anchor=west] {};
       \filldraw[black] (0.3,1.3) circle (0.9pt) node[anchor=west] {};
       \filldraw[black] (1.3,1.3) circle (0.9pt) node[anchor=west] {};
       \filldraw[black] (1.3,0.3) circle (0.4pt) node[anchor=west] {};

       \path (0,0) node [below]{$v_5$};
       \path (0,1) node [left]{$v_6$};
       \path (1,1) node [right]{$v_7$};
       \path (1,0) node [below]{$v_8$};
       \path (0.3,0.3) node [below]{$v_1$};
       \path (0.3,1.3) node [left]{$v_2$};
       \path (1.3,1.3) node [right]{$v_3$};
       \path (1.3,0.3) node [below]{$v_4$};

       \path (1.5,0.3) node [right]{$y$};
       \path (-0.25,-0.2) node [below]{$x$};
       \path (0.3,1.5) node [above]{$z$};
\end{tikzpicture}
\caption{An example of $2$-signature code of length $3$}
\label{fig.1}
%\end{center}
\end{figure}
Then the set of vectors $\C$ corresponding to $\Hh_0$, that is, $\C=\{(0,0,0),(0,0,1),(0,1,1),(1,0,1),(1,1,0)\}$, is a $(3,5,2)$ $2$-signature code. {\color{black}The matrix representation of $\C$ is presented below.
\begin{equation*}
\C=\begin{bmatrix}
0 & 0 & 0 & 1 & 1  \\
0 & 0 & 1 & 0 & 1  \\
0 & 1 & 1 & 1 & 0
\end{bmatrix}.
\end{equation*}
}
\end{example}

{\color{black}
In the application of an $(n,M,2)$ $t$-signature code $\C$ to multimedia fingerprinting, suppose that $\C_1$ corresponds to the set of fingerprints distributed to a coalition set $J$. As introduced in Section \ref{sec:1}, once a forged copy produced by the coalition $J$ is confiscated, the dealer can calculate a result $\vec{r}$. By (\ref{$r_k$}), (\ref{$r$}) and (\ref{open convex polytope}), $\vec{r}$ is a point in $\Pp(\C_1)$. Then we can trace back to $\C_1$, or equivalently the coalition set $J$, by checking that {\color{black}to which open convex polytope, formed by a subset of $\C$ with size no more than $t$, $\vec{r}$ belongs}. The computational complexity of the tracing algorithm is $O(nM^t)$.
}

%by checking that the open convex polytope of which subset of $\C$ with size no more than $t$ contains $\vec{r}$. 

%by checking that {\color{black}to which open convex polytope, generated by a subset of $\C$ with size no more than $t$, $\vec{r}$ belongs}.

\subsection{Known results}
\label{sec2-3}

Let $A(n,t)$ denote the maximum size of a $t$-signature code of length $n$, and $A(n,w,t)$ denote the maximum size of a $t$-signature code of length $n$ and constant-weight $w$. A $t$-signature code of length $n$ is called \textit{optimal} if it has size $A(n,t)$, and a $t$-signature code of length $n$ and constant-weight $w$ is called \textit{optimal} if it has size $A(n,w,t)$. Denote the largest asymptotic code rate of $t$-signature codes as
\begin{equation*}
  R(t)={\color{black}\limsup_{n\to\infty}}\frac{\log_2 A(n,t)}{n}.
\end{equation*}

In \cite{EFKL}, Egorova et al. found that a binary code with any $2t$ codewords linearly independent is a $t$-signature code. Based on this observation, they constructed signature codes from binary \textit{Goppa codes} in coding theory and obtained a lower bound on $A(n,t)$.

{\color{black}
\begin{theorem}\label{lower bound}{\rm(\cite{EFKL})}
{\color{black}
\begin{equation*}
  A(n,t)\ge 2^{\lfloor \frac{n}{t}\rfloor}.
\end{equation*}
}
\end{theorem}
}

This immediately implies

\begin{corollary}\label{coro:known rate of sig}
Let $t\ge 2$ be an integer. Then $R(t)\ge 1/t.$
\end{corollary}

In the literature, the exact values of $A(n,t)$ and $R(t)$ are generally unknown except the lower bounds in Theorem \ref{lower bound} and Corollary \ref{coro:known rate of sig}.
However, the lower bound on $A(n,t)$ in Theorem \ref{lower bound} is not tight in general since for instance it shows $A(3,2)\ge 2$, while Example \ref{3-dim cube} implies $A(3,2)\ge 5$. The main purpose of this paper is to explore bounds {\color{black}on} $A(n,t)$ and $A(n,w,t)$ by using combinatorial methods, and to provide constructions for $t$-signature codes which are equipped with efficient tracing algorithms.

\subsection{Relationships with $B_t$ codes}
\label{sec1,subsec2:related works}
{\color{black}
In this subsection, we build relationships between signature codes and $B_t$ codes.
}
The investigation of $B_t$ code was motivated by \textit{$B_t$-sequence} introduced by Erd\H os and Tur\'an \cite{ET}. $B_t$ codes for $t=2$ were first studied by Lindstr\"om in \cite{L69} and later in \cite{L72} under the name of \textit{$B_2$-sequences of vectors}. In recent decades, $B_t$ codes were investigated due to their applications to {\color{black}the} \textit{multiple access adder channel}. The interested reader is referred to \cite{D,DR81,DR83,DR81-1,GFM} for more details.

{\color{black}
\begin{definition}\label{def:B-code}
{\color{black}An $(n,M,q)$ code $\C$ is a $B_t$ code, or $(n,M,q)$ $B_t$ code, if for any two distinct multi-subsets $\widehat{\C}_1=\{\vec{c}_{j_1},\ldots,\vec{c}_{j_t}\}$ and $\widehat{\C}_2=\{\vec{c}_{k_1},\ldots,\vec{c}_{k_t}\}$ of $\C$, we have
\begin{equation*}
  \sum_{h=1}^t\vec{c}_{j_h}\neq \sum_{l=1}^t\vec{c}_{k_l}.
\end{equation*}
}
\end{definition}
}

From Definition \ref{def:B-code}, it is obvious that when $q=2$ and $t=2$, an $(n,M,2)$ code $\C$ is $B_2$ code if and only if $\vec{c}_i+\vec{c}_j\neq\vec{c}_k+\vec{c}_s$ for any distinct $\vec{c}_i,\vec{c}_j,\vec{c}_k,\vec{c}_s\in \C$. We have the following observation on the relationship between signature codes and $B_t$ codes.

{\color{black}
\begin{lemma}\label{sig. code and B code}
{\color{black}
Let $\C$ be an $(n,M,2)$ code and $t\ge 2$ be an integer. If $\C$ is a $t$-signature code, then $\C$ is a $B_t$ code.
}
\end{lemma}
\begin{IEEEproof} Let $\C=\{\vec{c}_1,\ldots,\vec{c}_M\}$ be an $(n,M,2)$ code. Assume that $\C$ is a $t$-signature code, but not a $B_t$ code. By Definition \ref{def:B-code}, there exist two distinct multi-subsets $\widehat{\C}_1=\{\vec{c}_{j_1},\ldots,\vec{c}_{j_t}\}$ and $\widehat{\C}_2=\{\vec{c}_{k_1},\ldots,\vec{c}_{k_t}\}$ of $\C$ such that $\sum_{h=1}^t\vec{c}_{j_h}=\sum_{l=1}^t\vec{c}_{k_l}$. By subtracting all the common codewords counted in the two summations, we can obtain a non-empty multi-subset $\widehat{\A}_1=\{\vec{c}_{j_1}',\ldots,\vec{c}_{j_m}'\}$ of $\widehat{\C}_1$ and a non-empty multi-subset $\widehat{\A}_2=\{\vec{c}_{k_1}',\ldots,\vec{c}_{k_m}'\}$ of $\widehat{\C}_2$ which satisfy $m\le t$, $\sum_{h=1}^m\vec{c}_{j_h}'=\sum_{l=1}^m\vec{c}_{k_l}'$ and the base sets of $\widehat{\A}_1,\widehat{\A}_2$ are distinct. Let $\lambda_{j_h}=\lambda_{k_l}=1/m$ for any $1\le h,l\le m$. Then we have $\sum_{h=1}^m\lambda_{j_h}=\sum_{l=1}^m\lambda_{k_l}=1$. Furthermore, we have $\sum_{h=1}^m\lambda_{j_h}\vec{c}_{j_h}'=\sum_{l=1}^m\lambda_{k_l}\vec{c}_{k_l}'$, a contradiction to the assumption that $\C$ is a $t$-signature code by Proposition \ref{def-general-equiv}. The lemma follows.
\end{IEEEproof}
}

The converse of Lemma~\ref{sig. code and B code} does not hold in general. However, {\color{black}when $t=2$, we have the following observation.}

{\color{black}
\begin{lemma}\label{2-sig equiv}
{\color{black}
Let $\C$ be an $(n,M,2)$ code. Then $\C$ is a $2$-signature code if and only if $\vec{c}_i+\vec{c}_j\neq\vec{c}_k+\vec{c}_s$ for any distinct $\vec{c}_i,\vec{c}_j,\vec{c}_k,\vec{c}_s\in\C$. In other words, $\C$ is a $2$-signature code if and only if it is a $B_2$ code.
}
\end{lemma}
}
\begin{IEEEproof}
The necessity is clear from Lemma \ref{sig. code and B code}. Now we show the sufficiency. Assume that for any distinct $\vec{c}_i,\vec{c}_j,\vec{c}_k,\vec{c}_s\in\C$, $\vec{c}_i+\vec{c}_j\neq\vec{c}_k+\vec{c}_s$, but $\C$ is not a $2$-signature code. Then according to Definition \ref{sig. code}, there exist two distinct subsets $\C_1,\C_2\subseteq\C$ with $1\le|\C_1|,|\C_2|\le 2$, and also exist $\lambda_j,\lambda_k'>0$ such that $\sum_{\vec{c}_j\in\C_1}\lambda_j\vec{c}_j=\sum_{\vec{c}_k\in\C_2}\lambda_k'\vec{c}_k$ where $\sum_{\vec{c}_j\in\C_1}\lambda_j=\sum_{\vec{c}_k\in\C_2}\lambda_k'=1$. We discuss the following cases.
\begin{enumerate}
  \item If $|\C_1|=1$ or $|\C_2|=1$, then we must have $\C_1=\C_2$, a contradiction to the assumption.
  \item If $|\C_1|=|\C_2|=2$ and $\C_1\cap \C_2\neq\emptyset$, then we can have $\C_1=\C_2$, a contradiction.
  \item If $|\C_1|=|\C_2|=2$ and $\C_1\cap \C_2=\emptyset$, without loss of generality, we may assume that $\C_1=\{\vec{c}_1,\vec{c}_2\}$ and $\C_2=\{\vec{c}_3,\vec{c}_4\}$. According to the assumption, there exist $\lambda_1,\lambda_2,\lambda_3',\lambda_4'>0$ such that $\lambda_1\vec{c}_1+\lambda_2\vec{c}_2=\lambda_3'\vec{c}_3+\lambda_4'\vec{c}_4$ where $\lambda_1+\lambda_2=\lambda_3'+\lambda_4'=1$. For any $1\le i\le n$, we discuss the following three subcases of $\{c_1(i),c_2(i)\}$.
      \begin{enumerate}
        \item[3.a)] If $\{c_1(i),c_2(i)\}=\{0\}$, then $\lambda_1c_1(i)+\lambda_2c_2(i)=0=\lambda_3'c_3(i)+\lambda_4'c_4(i)$, which implies $c_3(i)=c_4(i)=0$ since $\lambda_3',\lambda_4'>0$.
        \item[3.b)] If $\{c_1(i),c_2(i)\}=\{1\}$, then $\lambda_1 c_1(i)+\lambda_2 c_2(i)=1=\lambda_3'c_3(i)+\lambda_4'c_4(i)$ since $\lambda_1+\lambda_2=1$, which implies $c_3(i)=c_4(i)=1$ since $\lambda_3',\lambda_4'>0$ and $\lambda_3'+\lambda_4'=1$.
        \item[3.c)] If $\{c_1(i),c_2(i)\}=\{0,1\}$, then $0<\lambda_1c_1(i)+\lambda_2c_2(i)=\lambda_3'c_3(i)+\lambda_4'c_4(i)<1$, which implies $\{c_3(i),c_4(i)\}=\{0,1\}$.
      \end{enumerate}
      Then for case 3), we have $\vec{c}_1+\vec{c}_2=\vec{c}_3+\vec{c}_4$, also a contradiction to the assumption since $\{\vec{c}_1,\vec{c}_2\}\cap\{\vec{c}_3,\vec{c}_4\}=\emptyset$.
\end{enumerate}

The conclusion follows.
\end{IEEEproof}

%{\color{blue}By Definition \ref{def:B-code} and Lemmas \ref{sig. code and B code} and \ref{2-sig equiv}, it is obvious that a $t$-signature code is a binary $B_t$ code for any integer $t\ge 2$, and the converse holds when $t=2$.}

%{\color{black}
%By Lemmas \ref{sig. code and B code} and \ref{2-sig equiv}, we illustrate the relationships between signature codes and $B_t$ codes as follows.
%
%\begin{itemize}
%  \item For any integer $t\ge 3$: (``$\Rightarrow$'' means ``imply'')
%  \begin{equation}\label{Eq:relationship-1}
%    (n,M,2)\ t\text{-signature\ code}\ \Rightarrow\ (n,M,2)\ B_t\ \text{code}
%  \end{equation}
%
%  \item For $t=2$: (``$\Leftrightarrow$'' means ``equivalent to'')
%  \begin{equation}\label{Eq:relationship-2}
%   (n,M,2)\ 2\text{-signature\ code}\ \Leftrightarrow\ (n,M,2)\ B_2\ \text{code}
%  \end{equation}
%\end{itemize}
%}

\section{Upper bounds {\color{black}on $t$-signature codes}}
\label{sec:2}

In this section, we investigate upper bounds on $A(n,t)$ {\color{black}and $R(t)$}. First,
we derive a general upper bound on $A(n,t)$ by connecting a $t$-signature code with a bipartite graph containing no cycles of length less than or equal to $2t$, {\color{black}which also implies an upper bound on $R(t)$.
} Second,
we give an improved upper bound for $R(t)$ based on the known results on $B_t$ codes.

{\color{black}
\subsection{General upper bounds on $t$-signature codes}
\label{subsec:2-2}
}
%In the preceding subsection, we obtained an {\color{black}upper bound on the largest asymptotic code rate of $t$-signature codes.} In this subsection we show a general upper bound {\color{black}on} $A(n,t)$ for any positive {\color{black}integers $n,t\ge 2$} by relating a $t$-signature code to a bipartite graph containing no cycles of length less than or equal to $2t$.

Denote $G=(V_1,V_2,E)$ as a {\it bipartite graph} where $V_1,V_2$ are two disjoint sets of vertices and $E$ is the set of edges each connecting one vertex in $V_1$ and the other in $V_2$. $G=(V_1,V_2,E)$ is called {\it complete} if every vertex in $V_1$ is connected to every vertex in $V_2$. A complete bipartite graph $G=(V_1,V_2,E)$ with $|V_1|=m_1$ and $|V_2|=m_2$ is denoted as $K_{m_1,m_2}$. A {\it cycle} $C_k$ of $G=(V_1,V_2,E)$ is a sequence of vertices $v_1v_2\cdots v_kv_1$, where $v_i$, $1\le i\le k$ are all distinct and $\{v_i,v_{i+1}\},\{v_k,v_1\}\in E$ for any $1\le i\le k-1$. The number of vertices in a cycle is called the {\it length} of this cycle, and the {\it girth} of a graph $G$ is the length of the shortest cycle in $G$. If a graph $G$ contains no cycle of length $k$, it is called {\it $C_k$-free}. A {\it matching} of $G$ is a subset of edges without common vertices, and a {\it perfect matching} is a matching which matches all vertices in $G$.

Let $\C$ be an $(n,M,2)$ code. For a partition of $[n]=I_1\cup I_2$ with $|I_1|=n_1$, $|I_2|=n_2$ and $n_1+n_2=n$, we define a bipartite graph corresponding to $\C$ as $G=(V_1,V_2,E)$ such that $V_1=\{0,1\}^{n_1}$, $V_2=\{0,1\}^{n_2}$ and for any $\vec{u}\in V_1$ and $\vec{v}\in V_2$, $\{\vec{u},\vec{v}\}\in E$ if and only if there exists a codeword $\vec{c}\in\C$ such that $\vec{c}|_{I_1}=\vec{u}$ and $\vec{c}|_{I_2}=\vec{v}$. Denote $e(G)$ as the number of edges in $G$. It is obvious that $|V_1|=2^{n_1}$, $|V_2|=2^{n_2}$ and $e(G)=|\C|$. Moreover, we have the following observation.

\begin{lemma}\label{sig. code and bipartite}
If $\C$ is a $t$-signature code, then for any partition of $[n]=I_1\cup I_2$, the girth of the corresponding bipartite graph $G=(V_1,V_2,E)$ is at least $2t+2$. %is $C_{2k}$-free for all $2\le k\le t$.
\end{lemma}
\begin{IEEEproof} {\color{black}Note that a bipartite graph contains no cycles of odd length.} To show $G=(V_1,V_2,E)$ has girth at least $2t+2$, it suffices to show that $G=(V_1,V_2,E)$ contains no cycles of length $2k$ for any $2\le k\le t$. Assume that $\C$ is a $t$-signature code, but $G=(V_1,V_2,E)$ contains a cycle of length $2k$ for some $2\le k\le t$. Without loss of generality, we may assume that the cycle $C_{2k}$ is $v_1v_2\ldots v_{2k-1}v_{2k}v_1${\color{black}, where $v_1,v_3,\ldots,v_{2k-1}\in V_1$ and $v_2,v_4,\ldots,v_{2k}\in V_2$}. Let $e_i=\{v_i,v_{i+1}\}$ for $1\le i\le 2k-1$, and $e_{2k}=\{v_{2k},v_1\}$. Then $\{e_i: 1\le i\le 2k,\ i\ \text{is\ odd}\}$ and $\{e_i: 1\le i\le 2k,\ i\ \text{is\ even}\}$ are two disjoint perfect matchings of the cycle $C_{2k}$. Recall that each edge of $G$ corresponds to a codeword of $\C$. Let $\vec{c}_i\in\C$ be the corresponding codeword of $e_i$ in $C_{2k}$ for $1\le i\le 2k$, and $\C_1=\{\vec{c}_i: 1\le i\le 2k,\ i\ \text{is\ odd}\}$, $\C_2=\{\vec{c}_i: 1\le i\le 2k,\ i\ \text{is\ even}\}$. Then $\C_1$ and $\C_2$ are two disjoint subsets of $\C$ such that $|\C_1|=|\C_2|=k${\color{black}, $\sum_{\vec{c}\in\C_1}\vec{c}|_{I_1}=\sum_{\vec{c}\in\C_2}\vec{c}|_{I_1}=\sum_{i \ \text{is odd}}v_i$ and $\sum_{\vec{c}\in\C_1}\vec{c}|_{I_2}=\sum_{\vec{c}\in\C_2}\vec{c}|_{I_2}=\sum_{i \ \text{is even}}v_i$. Therefore, we have} $\sum_{\vec{c}\in\C_1}\vec{c}=\sum_{\vec{c}\in\C_2}\vec{c}$, a contradiction to Lemma \ref{sig. code and B code} since $\C$ is a $t$-signature code. The conclusion follows.
\end{IEEEproof}

The following result on bipartite {\color{black}graphs} was shown in \cite{NV}.

{\color{black}
\begin{lemma}{\rm(\cite{NV})}\label{Thm:upper of bi-graph}
Let $G=(V_1,V_2,E)$ be a bipartite graph with $|V_1|=u$ and $|V_2|=v$. If $G$ is $C_{2t}$-free, then
\begin{equation*}
  e(G)\le \left\{\begin{array}{ll}
  (2t-3)(v^{\frac{1}{2}}u^{\frac{t+2}{2t}}+u+v), & t\ \text{is\ even}, \\
  (2t-3)((uv)^{\frac{t+1}{2t}}+u+v), & t\ \text{is\ odd}.
\end{array}\right.
\end{equation*}
\end{lemma}
}

By Lemmas \ref{sig. code and bipartite} and \ref{Thm:upper of bi-graph}, we obtain

{\color{black}
\begin{theorem}\label{Thm:general upper of sig}
Let $n,t\ge 2$ be integers. Then
\begin{equation}\label{Eq:general uppder of A(n,t)}
  A(n,t)\le\left\{\begin{array}{ll}
  (2t-3)\big(2^{\frac{(t+2)n}{2t+2}+1}+2^{\frac{tn}{2t+2}}\big), & t\ \text{is\ even}, \\
  (2t-3)\big(2^{\frac{(t+1)n}{2t}}+2^{\lceil\frac{n}{2}\rceil}+
             2^{\lfloor\frac{n}{2}\rfloor}\big), & t\ \text{is\ odd}.
\end{array}\right.
\end{equation}
\end{theorem}
\begin{IEEEproof}
{\color{black}
Let $\C$ be a $t$-signature code of length $n$ and size $M=A(n,t)$. By Lemma \ref{sig. code and bipartite}, for any partition of $[n]=I_1\cup I_2$ with $|I_1|=n_1$, $|I_2|=n_2$ and $n_1+n_2=n$, the corresponding bipartite graph $G=(V_1,V_2,E)$ with $|V_1|=2^{n_1}$, $|V_2|=2^{n_2}$ and $e(G)=A(n,t)$ is $C_{2k}$-free for any $2\le k\le t$. Then by {\color{black}Lemma} \ref{Thm:upper of bi-graph}, we have that
\begin{equation*}
  A(n,t)\le \left\{\begin{array}{ll}
  (2t-3)(2^{\frac{t+2}{2t}n_1+\frac{n_2}{2}}+2^{n_1}+2^{n_2}), & t\ \text{is\ even}, \\
  (2t-3)(2^{\frac{t+1}{2t}n}+2^{n_1}+2^{n_2}), & t\ \text{is\ odd},
\end{array}\right.
\end{equation*}
holds for any integers $n_1,n_2$ such that $n_1,n_2\ge 0$ and $n_1+n_2=n$. To make the upper bound on $A(n,t)$ as small as possible, we take
\begin{equation*}%\label{n1-n2}
  (n_1,n_2)=\left\{\begin{array}{ll}
  (\frac{tn}{2t+2},\frac{(t+2)n}{2t+2}), & t\ \text{is\ even}, \\
  (\lceil\frac{n}{2}\rceil,\lfloor\frac{n}{2}\rfloor), & t\ \text{is\ odd},
\end{array}\right.
\end{equation*}
and then (\ref{Eq:general uppder of A(n,t)}) follows.
}
\end{IEEEproof}
}

We mention that
the above approach of upper bounding the size of a code by exploring its connection with graphs without small cycles was used in~\cite{CFJLM,Cheng2015,DPSV} as well. When $t=2$, Theorem \ref{Thm:general upper of sig} shows that $A(n,2)\le 2\cdot 2^{\frac{2n}{3}}+2^{\frac{n}{3}}$. A better upper bound for $A(n,2)$ could be obtained by using Roman's bound \cite{R75} on bipartite graphs.

{\color{black}
\begin{lemma}{\rm(\cite{R75})}\label{Thm:upper of bi-graph-2}
Let $G=(V_1,V_2,E)$ be a bipartite graph with $|V_1|=u$, $|V_2|=v$, and let $p$ be an integer such that $p\ge s-1$. If G contains no $K_{s,m}$, then
\begin{equation*}
  e(G)\le \frac{m-1}{\binom{p}{s-1}}\binom{u}{s}+v\frac{(p+1)(s-1)}{s}.
\end{equation*}
\end{lemma}
}

By Lemmas \ref{sig. code and bipartite} and \ref{Thm:upper of bi-graph-2}, we have

{\color{black}
\begin{theorem}\label{Thm:upper of 2-sig}
Let $n\ge 2$ be an integer. Then
\begin{equation}\label{Eq:upper of A(n,2)}
  A(n,2)\le\left\{\begin{array}{ll}
  2^{\lfloor\frac{2n}{3}\rfloor}+2^{\lceil\frac{n}{3}\rceil}(2^{\lceil\frac{n}{3}\rceil}-1)/2, & n\equiv 2\pmod 3, \\
  2^{\lceil\frac{2n}{3}\rceil}+2^{\lfloor\frac{n}{3}\rfloor}(2^{\lfloor\frac{n}{3}\rfloor}-1)/2, & \text{otherwise}.
\end{array}\right.
  %A(n,2)\le 2^{\frac{2n}{3}}+2^{\frac{n}{3}}(2^{\frac{n}{3}}-1)/2.
\end{equation}
\end{theorem}
\begin{IEEEproof}
{\color{black}
Let $\C$ be a $2$-signature code of length $n$ and size $M=A(n,2)$. Similarly, according to Lemma \ref{sig. code and bipartite}, for any partition of $[n]=I_1\cup I_2$ with $|I_1|=n_1$, $|I_2|=n_2$ and $n_1+n_2=n$, the corresponding bipartite graph $G=(V_1,V_2,E)$ with $|V_1|=2^{n_1}$, $|V_2|=2^{n_2}$ and $e(G)=A(n,2)$ is $C_4$-free. Note that $C_4\cong K_{2,2}$. Then by choosing $s=m=2$ and $p=1$ in Lemma \ref{Thm:upper of bi-graph-2}, we have that
\begin{equation*}%\label{Eq:upper of bi-graph}
  A(n,2)\le 2^{n_2}+2^{n_1}(2^{n_1}-1)/2
\end{equation*}
holds for any integers $n_1,n_2$ such that $n_1,n_2\ge 0$ and $n_1+n_2=n$. To make the upper bound on $A(n,2)$ as small as possible, we take
\begin{equation*}%\label{improved n1-n2}
  (n_1,n_2)=\left\{\begin{array}{ll}
  (\lceil\frac{n}{3}\rceil,\lfloor\frac{2n}{3}\rfloor), & n\equiv 2\pmod 3, \\
  (\lfloor\frac{n}{3}\rfloor,\lceil\frac{2n}{3}\rceil), & \text{otherwise}.
\end{array}\right.
\end{equation*}
Then (\ref{Eq:upper of A(n,2)}) follows.
}
\end{IEEEproof}
}

It can be checked that the upper bound on $A(n,2)$ in Theorem \ref{Thm:upper of 2-sig} is {\color{black}tighter} than that in Theorem \ref{Thm:general upper of sig} for the case $t=2$. {\color{black}Indeed, for instance Theorem \ref{Thm:general upper of sig} tells $A(3,2)\le 10$ while Theorem \ref{Thm:upper of 2-sig} shows $A(3,2)\le 5$.} The following result shows that the upper bound on $A(n,2)$ in Theorem \ref{Thm:upper of 2-sig} is tight for certain cases.

\begin{corollary}\label{$A(2,2)$ and $A(3,2)$}
{\color{black}
$A(2,2)=3$, $A(3,2)=5$ and $A(4,2)=7$.
}
\end{corollary}

\begin{IEEEproof}
For $n=2$, it is easy to check that $\C=\{(1,0),(0,1),(1,1)\}$ is a $2$-signature code. Moreover, Theorem \ref{Thm:upper of 2-sig} shows that $A(2,2)\le 3$, resulting $A(2,2)=3$.

For $n=3$, Theorem \ref{Thm:upper of 2-sig} shows that $A(3,2)\le 5$ which, together with Example \ref{3-dim cube}, yields $A(3,2)=5$.

For $n=4$, we have that $\C=\{(1,0,0,0),(0,1,0,0),(0,0,1,0),(1,1,0,0),(1,1,1,0),(0,0,0,1),(1,1,1,1)\}$ is a $2$-signature code, implying $A(4,2)\ge 7$. Notice that Theorem \ref{Thm:upper of 2-sig} only tells that $A(4,2)\le 9$. However, we can show $A(4,2)\le 7$ whose proof is deferred to Appendix, and thus we have $A(4,2)=7$.
\end{IEEEproof}

From Theorem \ref{Thm:general upper of sig}, we can obtain an upper bound on $R(t)$ for any integer $t\ge 2$. Here, we remark that although the upper bound on $A(n,2)$ in Theorem \ref{Thm:upper of 2-sig} is tighter than that in Theorem \ref{Thm:general upper of sig} for $t=2$, they imply the same upper bound on the largest asymptotic code rate $R(2)\le 2/3$. In general, by Theorem \ref{Thm:general upper of sig}, we have

\begin{corollary}\label{coro:rate of sig-1}
Let $t\ge 2$ be an integer. Then
\begin{equation*}
  R(t)\le \left\{\begin{array}{ll}
  \frac{1}{2}+\frac{1}{2t+2}, & t\ \text{is\ even}, \\
  \frac{1}{2}+\frac{1}{2t}, & t\ \text{is\ odd}.
\end{array}\right.
\end{equation*}
\end{corollary}

\subsection{Asymptotic upper bounds on $t$-signature codes}
\label{subsec:2-1}

{\color{black}In this subsection, we provide an improved upper bound on $R(t)$} based on Lemma \ref{sig. code and B code} and the known results on $B_t$ codes.

Denote $B(n,t)$ as the maximum size of a binary $B_t$ code of length $n$. Define the largest asymptotic code rate of binary $B_t$ codes as
\begin{equation*}
  R_B(t)={\color{black}\limsup_{n\to\infty}}\frac{\log_2 B(n,t)}{n}.
\end{equation*}
Denote $H_t=-\sum_{k=0}^t\binom{t}{k}2^{-t}\log_2\big(\binom{t}{k}2^{-t}\big)$ and $h_t=\log_2(t+1)+t/(t+1)$. The best known {\color{black}upper bounds on} $R_B(t)$ are as follows.%results of

{\color{black}
\begin{lemma}\label{known B code}%{\rm(\cite{CLZ,D,DR81,DR81-1,DR83})}
{\color{black}Let $t\ge 2$ be an integer. Then}
\begin{enumerate}
  \item {\color{black}{\rm(\cite{CLZ,DR81-1})}} {\color{black}$R_B(2)\le 0.5753$ and $R_B(4)\le 0.4451$.}

  \item {\color{black}{\rm(\cite{DR81,DR83})}} {\color{black}$R_B(2t-1)\le
    \left\{\begin{array}{l}
      (t/H_t+(t-1)/h_t)^{-1},\ 2\le t\le 5, \\
      H_{2t-1}/(2t-1),\ t\ge 6.
    \end{array}\right.$}

  \item {\color{black}{\rm(\cite{DR81,DR83})}} {\color{black}$R_B(2t)\le
    \left\{\begin{array}{l}
     (t/H_t+t/h_t)^{-1},\ 3\le t\le 5, \\
     H_{2t}/(2t),\ t\ge 6.
    \end{array}\right.$}

  \item {\color{black}{\rm(\cite{D})}} {\color{black}For sufficiently large $t$, $R_B(t)\le (\log_2t+\log_2(\pi e/2))/(2t)+O(1/t^3)$.}
\end{enumerate}
\end{lemma}
}

It is obvious from
{\color{black}Lemmas \ref{sig. code and B code} and \ref{2-sig equiv} that $R(t)\le R_B(t)$ for any integer $t\ge 2$ and the equality holds for $t=2$.
}
Combining this with Lemma \ref{known B code}, we obtain

{\color{black}
\begin{theorem}\label{asymptotic upper bound}
{\color{black}Let $t\ge 2$ be an integer. Then
\begin{enumerate}
  \item $R(2)\le 0.5753$ and $R(4)\le 0.4451$.
  \item $R(2t-1)\le\left\{\begin{array}{l}
  (t/H_t+(t-1)/h_t)^{-1},\ 2\le t\le 5, \\
  H_{2t-1}/(2t-1),\ t\ge 6.
  \end{array}\right.$
  \item $R(2t)\le\left\{\begin{array}{l}
  (t/H_t+t/h_t)^{-1},\ 3\le t\le 5, \\
  H_{2t}/(2t),\ t\ge 6.
  \end{array}\right.$
  \item For sufficiently large integer $t$, $R(t)\le(\log_2t+\log_2(\pi e/2))/(2t)+O(1/t^3)$.
\end{enumerate}
}
\end{theorem}
}

{\color{black}
We remark that the upper bound of $R(t)$ obtained in Theorem \ref{asymptotic upper bound} is tighter than that obtained in Corollary \ref{coro:rate of sig-1}.} For example, Corollary \ref{coro:rate of sig-1} implies $R(2)\le 2/3$ and $R(4)\le 0.6$, while Theorem \ref{asymptotic upper bound} shows that $R(2)\le 0.5753$ and $R(4)\le 0.4451$. %For any sufficiently large integer $t$, Corollary \ref{coro:rate of sig-1} implies $R(t)\le 1/2+O(1/t)$, while Theorem \ref{asymptotic upper bound} shows that $R(t)\le(\log_2t+\log_2(\pi e/2))/(2t)+O(1/t^3)<1/2+O(1/t)$. %A comparison between the upper bounds of $R(t)$ in Corollary \ref{coro:rate of sig-1} and Theorem \ref{asymptotic upper bound} for $3\le t\le 9$ is shown in Figure \ref{fig:comparision on asymptotic sig}.
%and $R(t)\le 1/2$ for any sufficiently large integer $t$, but Theorem \ref{asymptotic upper bound} shows that $R(2)\le 0.5753$ and $R(t)\le(\log_2t+\log_2(\pi e/2))/(2t)+O(1/t^3)<1/2$ for any sufficiently large integer $t$.

\section{{\color{black}Optimal $2$-signature codes with constant-weights $2$ and $3$}}
\label{sec:3}

In this section, we study the combinatorial properties of $2$-signature codes of length $n$ and constant-weights $2$ and $3$, respectively. Accordingly, {\color{black}bounds on} $A(n,2,2)$ and $A(n,3,2)$ are obtained{\color{black}, and optimal $2$-signature codes of length $n$ and constant-weights $2$ and $3$ for infinite values of $n$ are constructed respectively.}

% {\color{brown}Here we would like to mention that constant-weight signature codes also have some practical implications. As stated in Section~\ref{sec:1}, the fingerprint $\vec{w}$ of
%  an authorized user is represented as
% $
% \vec{w}=\sum_{i=1}^n c(i)\vec{f}_i,
% $
% where
% $\vec{f}_1,\ldots,\vec{f}_n\in\mathbb{R}^m$ are noise-like orthonormal signals and $\vec{c}=(c(1),\ldots,c(n))$ is a codeword.
% If the signature code is constant-weight, then each user's fingerprint is represented under $\{\vec{f}_i\}$
% with the same sparsity, which makes the study of constant-weight codes meaningful as well.}d

\subsection{Optimal $2$-signature codes with constant-weight $2$}
\label{subsec:3-1}

Let $\C$ be a binary code of length $n$ and constant-weight $2$. Let $G=(V,E)$ be the corresponding graph of $\C$ where $V=[n]$ and for any distinct $i,j\in [n]$, $\{i,j\}\in E$ if and only if there exists a codeword $\vec{c}\in\C$ such that $\mathrm{supp}(\vec{c})=\{i,j\}$. Then we have the following observation.
%\begin{equation*}
%c_s=\left\{\begin{array}{l}
%1,\  s\in\{i, j\},  \\
%0,\ \text{otherwise}.
%\end{array}\right.
%\end{equation*} Then we have the following observation.

{\color{black}
\begin{lemma}\label{2-sig code and graph}
{\color{black}
$\C$ is a $2$-signature code of constant-weight $2$ if and only if its corresponding graph $G$ is $C_4$-free.}
\end{lemma}
}
\begin{IEEEproof} We first show the necessity. If there exist four vertices $i,j,k,s\in [n]$ which form a cycle $ijksi$ in graph $G$, then there must exist four distinct codewords $\vec{c}_1,\vec{c}_2,\vec{c}_3,\vec{c}_4\in\C$ such that $\mathrm{supp}(\vec{c}_1)=\{i,j\}$, $\mathrm{supp}(\vec{c}_2)=\{j,k\}$, $\mathrm{supp}(\vec{c}_3)=\{k,s\}$ and $\mathrm{supp}(\vec{c}_4)=\{s,i\}$. It is easy to verify that $\vec{c}_1+\vec{c}_3=\vec{c}_2+\vec{c}_4$, a contradiction to Lemma \ref{sig. code and B code}.

On the other hand, assume that the corresponding graph $G$ of $\C$ is $C_4$-free, but $\C$ is not a $2$-signature code. By Lemma \ref{sig. code and B code}, there must exist four distinct codewords $\vec{c}_i,\vec{c}_j,\vec{c}_k,\vec{c}_s\in\C$ such that $\vec{c}_i+\vec{c}_j=\vec{c}_k+\vec{c}_s$. If $\mathrm{supp}(\vec{c}_i)\cap\mathrm{supp}(\vec{c}_j)\neq\emptyset$, then we have $\{\vec{c}_i,\vec{c}_j\}=\{\vec{c}_k,\vec{c}_s\}$, a contradiction to the assumption. Thus there must be four elements in $\mathrm{supp}(\vec{c}_i)\cup \mathrm{supp}(\vec{c}_j)=\mathrm{supp}(\vec{c}_k)\cup\mathrm{supp}(\vec{c}_s)$ which form a cycle of length $4$ in $G$, a contradiction to the assumption. The lemma follows.
\end{IEEEproof}

Lemma \ref{2-sig code and graph} shows that if there exists a $2$-signature code with constant-weight $2$, then we can construct a $C_4$-free graph, and conversely, given a $C_4$-free graph, we can obtain a $2$-signature code with constant-weight $2$. The following example illustrates this relationship.

{\color{black}
\begin{example}\label{c4-free-fraph}
{\color{black}
Consider the $C_4$-free graph on $5$ vertices depicted in Figure \ref{fig:cycle-four-free}.
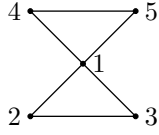
\begin{figure}
%\begin{center}
\begin{tikzpicture}[scale=0.7]
       \coordinate (1) at (0, 0);
       \coordinate (2) at (-1,-1);
       \coordinate (3) at (1,-1);
       \coordinate (4) at (-1,1);
       \coordinate (5) at (1,1);
%       \coordinate (6) at (-1.4,-0.8);
%       \coordinate (7) at (1.4,-0.8);
       \draw[line width=0.5pt] (1) -- (2);
       \draw[line width=0.5pt] (1) -- (3);
       \draw[line width=0.5pt] (1) -- (4);
       \draw[line width=0.5pt] (1) -- (5);
       \draw[line width=0.5pt] (2) -- (3);
       \draw[line width=0.5pt] (4) -- (5);
       \filldraw[black] (0,0) circle (1.3pt) node[anchor=west] {};
       \filldraw[black] (-1,-1) circle (1.3pt) node[anchor=west] {};
       \filldraw[black] (1,-1) circle (1.3pt) node[anchor=west] {};
       \filldraw[black] (-1,1) circle (1.3pt) node[anchor=west] {};
       \filldraw[black] (1,1) circle (1.3pt) node[anchor=west] {};
       \path (0,0) node [right]{$1$};
       \path (-1,-1) node [left]{$2$};
       \path (1,-1) node [right]{$3$};
       \path (-1,1) node [left]{$4$};
       \path (1,1) node [right]{$5$};
\end{tikzpicture}
\caption{The $C_4$-free graph in Example \ref{c4-free-fraph}}
%\captionof{figure}{$C_4$-free graph in Example \ref{c4-free-fraph}}
\label{fig:cycle-four-free}
%\end{center}
\end{figure}
Then according to Lemma \ref{2-sig code and graph}, we can obtain a $2$-signature code $\C$ of length $5$ and constant-weight $2$ as follows.% by defining a bijection $\sigma:\ \{v_1\}$
\begin{equation*}
  \C=\begin{bmatrix}
   1 & 1 & 0 & 1 & 1 & 0 \\
   1 & 0 & 1 & 0 & 0 & 0 \\
   0 & 1 & 1 & 0 & 0 & 0 \\
   0 & 0 & 0 & 1 & 0 & 1 \\
   0 & 0 & 0 & 0 & 1 & 1
  \end{bmatrix}.
\end{equation*}
}
\end{example}
}

Let $g(n;C_4)$ denote the maximum number of edges in a $C_4$-free graph with $n$ vertices. Then by Lemma \ref{2-sig code and graph}, we have $A(n,2,2)=g(n;C_4)$. It was proved in \cite{B,ERS} that {\color{black}$g(n;C_4)$ is asymptotically equal to $(1/2+o(1))n^{3/2}$.} In \cite{F83,F96,FS}, it was shown that if $n=m^2+m+1$ and $m$ is a power of $2$ or a prime power greater than $13$, then $g(n;C_4)=m(m+1)^2/2$ and a $C_4$-free graph with $n$ vertices and $m(m+1)^2/2$ edges could be constructed from the \textit{polarity of a projective plane of order $m$}, see \cite{ERS,F83,F96,FS} for more details. For any integer $2\le n\le 21$, the exact value of $g(n;C_4)$ and the corresponding $C_4$-free graph with $n$ vertices and $g(n;C_4)$ edges were presented in \cite{CFS}. Based on these known results for $C_4$-free graphs and Lemma \ref{2-sig code and graph}, we can obtain the following results for $2$-signature codes with constant-weight $2$.
%that {\color{black}$g(n;C_4)\le m(m+1)^2/2$ for any integer $n$ such that $n=m^2+m+1$ and $m\neq 1,7,9,11,13$. Moreover,
%The exact value of $g(n;C_4)$ is known for all values of $n$ of the form $n=m^2+m+1$, where $m$ is a power of $2$ \cite{F83}, or a prime power exceeding $13$ \cite{F94}, and it is equal to $\frac{m(m+1)^2}{2}$ The extremal graphs for these values of $n$ are known, and were constructed in

%{\color{black}Based on these known results for $C_4$-free graphs and Lemma \ref{2-sig code and graph}, we can obtain the following results for $2$-signature codes with constant-weight $2$.}

\begin{theorem}\label{$A(n,2,2)$}
Let $n\ge 2$ be an integer.
\begin{enumerate}
  \item {\color{black}$A(n,2,2)$ is asymptotically equal to $(1/2+o(1))n^{3/2}$.}%If $n$ is sufficiently large, then $A(n,2,2)=(1/2+o(1))n^{3/2}$.

  \item If $n=m^2+m+1$ where $m$ is a power of $2$ or $m>13$ is a prime power, then $A(n,2,2)=m(m+1)^2/2$ and an optimal $2$-signature code of length $n$ and constant-weight $2$ can be constructed from the polarity of a projective plane of order $m$.

  \item For $2\le n\le 21$, the exact values of $A(n,2,2)$ are listed in Table \ref{exact values}, and optimal $2$-signature codes of length $n$ and constant-weight $2$ can be constructed from the corresponding $C_4$-free graphs presented in \cite{CFS}.
  \begin{table*}
  \centering
\caption{The values of $A(n,2,2)$ for $2\le n\le 21${\color{black}~(\cite{CFS})}}
  \label{exact values}
  \begin{tabular}{c|cccccccccc}
    \hline
    $n$ & $2$ & $3$ & $4$ & $5$ & $6$ & $7$ & $8$ & $9$ & $10$ & $11$ \\
    \hline
    $A(n,2,2)$ & $1$ & $3$ & $4$ & $6$ & $7$ & $9$ & $11$ & $13$ & $16$ & $18$ \\
    \hline\hline
    $n$ & $12$ & $13$ & $14$ & $15$ & $16$ & $17$ & $18$ & $19$ & $20$ & $21$ \\
    \hline
    $A(n,2,2)$ & $21$ & $24$ & $27$ & $30$ & $33$ & $36$ & $39$ & $42$ & $46$ & $50$ \\
    \hline
  \end{tabular}
  %\\
  %\vspace{20pt}
  %\hrulefill
  %\vspace*{4pt}
  \end{table*}
\end{enumerate}
\end{theorem}

{\color{black}We remark that Theorem \ref{$A(n,2,2)$} is in fact a consequence of the combination of Lemma \ref{2-sig code and graph} and the known results for $C_4$-free graphs outlined above.} Besides, the $C_4$-free graph depicted in Figure \ref{fig:cycle-four-free} was presented in \cite{CFS}. By Theorem \ref{$A(n,2,2)$}, the obtained $(5,6,2)$ $2$-signature code $\C$ with constant-weight $2$ in Example \ref{c4-free-fraph} is optimal.

\subsection{Optimal $2$-signature codes with constant-weight $3$}
\label{subsec:3-2}

In this subsection, we investigate the {\color{black}combinatorial} properties of $2$-signature codes of constant-weight $3$ by means of Frankl and F\"uredi's method in \cite{FF}, where they studied binary codes using techniques in extremal set theory.
%In, Frankl and F\"uredi studied binary codes of constant weight $3$ from the point-view of. In this subsection, we investigate the extremal property of $2$-signature codes of constant weight $3$ by their method.

Let $\C$ be a binary code of length $n$ and constant-weight $3$. Notice that $\C$ can be uniquely described as a family $\F\subseteq \binom{[n]}{3}$ where $\F=\{\mathrm{supp}(\vec{c}): \vec{c}\in\C\}$. For any $A\in\binom{[n]}{2}$, denote
\begin{equation}\label{$T(A)$}
  T(A)=\{z\in[n]: A\cup\{z\}\in\F\}.
\end{equation}
Then we have the following observation.

{\color{black}
\begin{lemma}\label{suff. and necess. of $2$-sig.}
{\color{black}$\C$ is a $2$-signature code of constant-weight $3$ if and only if $|T(A)\cap T(A')|\le 1$ for any distinct $A,A'\in\binom{[n]}{2}$.}
\end{lemma}
}
\begin{IEEEproof}
To show the necessity, assume that there exist distinct $A,A'\in\binom{[n]}{2}$ and also distinct elements $z,z'\in [n]$ such that $\{z,z'\}\subseteq T(A)\cap T(A')$. Then $A\cup\{z\}, A\cup\{z'\}, A'\cup\{z\}, A'\cup\{z'\}$ are four distinct members of $\F$. Suppose that their corresponding codewords in $\C$ are $\vec{c}_1, \vec{c}_2, \vec{c}_3, \vec{c}_4$, respectively. Then we have $\vec{c}_1+\vec{c}_4=\vec{c}_2+\vec{c}_3$, a contradiction to Lemma \ref{sig. code and B code}.

Next, we show the sufficiency. Assume that $\C$ satisfies that $|T(A)\cap T(A')|\le 1$ for any distinct $A,A'\in\binom{[n]}{2}$, but is not a $2$-signature code. Then according to Lemma \ref{sig. code and B code}, there exist distinct codewords $\vec{c}_i,\vec{c}_j,\vec{c}_k,\vec{c}_s\in\C$ such that $\vec{c}_i+\vec{c}_j=\vec{c}_k+\vec{c}_s$. Let $B_m=\mathrm{supp}(\vec{c}_m)$, $m\in\{i,j,k,s\}$ which are four distinct members of $\F$. Suppose that $B_i=\{a,b,c\}$ and $B_j=\{x,y,z\}$. If $|B_i\cap B_j|=2$, then we must have $\{B_i,B_j\}=\{B_k,B_s\}$ which implies $\{\vec{c}_i,\vec{c}_j\}=\{\vec{c}_k,\vec{c}_s\}$, a contradiction to the assumption. So, we have $|B_i\cap B_j|\le 1$. Now we discuss the following two cases.
\begin{enumerate}
  \item If $|B_i\cap B_j|=1$, assume that $c=x$. Then $c\in B_k\cap B_s$. Recall that $\vec{c}_i+\vec{c}_j=\vec{c}_k+\vec{c}_s$ and $\{B_i,B_j\}\cap\{B_k,B_s\}=\emptyset$. Thus one of $a$ and $b$ should be in $B_k$ and the other in $B_s$, and so as $y$ and $z$. Without loss of generality, we may assume that $B_k=\{a,y,c\}$ and $B_s=\{b,z,c\}$. Let $A=\{a,c\}$ and $A'=\{z,c\}$. Then $\{b,y\}\in\binom{T(A)}{2}\cap\binom{T(A')}{2}$, a contradiction to that $|T(A)\cap T(A')|\le 1$.
  \item If $|B_i\cap B_j|=0$, then $B_k\cap B_s=\emptyset$. Without loss of generality, we may assume that $B_k=\{a,b,x\}$ and $B_s=\{c,y,z\}$. Let $A=\{a,b\}$ and $A'=\{y,z\}$. Then $\{c,x\}\in\binom{T(A)}{2}\cap\binom{T(A')}{2}$, also a contradiction.
\end{enumerate}

The conclusion then follows.
\end{IEEEproof}

Based on Lemma \ref{suff. and necess. of $2$-sig.}, we shall derive an upper bound for $A(n,3,2)$ which is the same as an upper bound for weakly union-free families given by Frankl and F\"uredi \cite{FF}. However, the bound cannot be obtained directly from the relationship between signature codes and weakly union-free families, which will be shown in Lemma \ref{weakly and $B_2$} afterwards.
%Here we expose the proof of this upper bound for self-containment. The relationship between signature codes and weakly union-free families will be shown in Lemma \ref{weakly and $B_2$} afterwards.

{\color{black}
\begin{theorem}\label{upper bound of $A(n,3,2)$}
{\color{black}$A(n,3,2)\le n(n-1)/3$ for any $n\ge 3$.}
\end{theorem}
}
\begin{IEEEproof}
Let $\C$ be an $(n,M,2)$ $2$-signature code of constant-weight $3$ with $M=A(n,3,2)$. According to (\ref{$T(A)$}), we have
\begin{equation}\label{sum2}
  3A(n,3,2)=\sum_{A\in \binom{[n]}{2}}|T(A)|
\end{equation}
since each $3$-elements set has $3$ subsets of size $2$ and each codeword in $\C$ is calculated three times in the right-hand side of (\ref{sum2}). Besides, Lemma \ref{suff. and necess. of $2$-sig.} tells that $\binom{T(A)}{2}\cap\binom{T(A')}{2}=\emptyset$ for any distinct $A,A'\in\binom{[n]}{2}$, implying
\begin{equation}\label{inequality2}
  \sum_{A\in \binom{[n]}{2}} \binom{|T(A)|}{2}\le\binom{n}{2}.
\end{equation}

Note that the maximum value of $\sum_{A\in \binom{[n]}{2}}|T(A)|$ under the condition (\ref{inequality2}) is achieved when $|T(A)|=2$ for all $A\in\binom{[n]}{2}$. Hence by (\ref{sum2}),
\begin{equation*}
  \begin{aligned}
     A(n,3,2) & \le 2\binom{n}{2}\big/ 3\\
      & =n(n-1)/3
   \end{aligned}
\end{equation*} and the theorem follows.
\end{IEEEproof}

Next, we show that the upper bound {\color{black}on} $A(n,3,2)$ in Theorem \ref{upper bound of $A(n,3,2)$} can be achieved for all values of $n$ with some exceptions by establishing the relationship between signature codes and weakly union-free families. A family $\F\subseteq 2^{[n]}$ is called {\it weakly union-free} if for any distinct $A,B,C,D\in\F$, $A\cup B\neq C\cup D$.

\begin{lemma}\label{weakly and $B_2$}
Let $\C$ be a binary code of length $n$ and $\F\subseteq 2^{[n]}$ be the corresponding family. If $\F$ is weakly union-free, then $\C$ is a $2$-signature code.
\end{lemma}
\begin{IEEEproof} Assume that $\F$ is weakly union-free, but $\C$ is not a $2$-signature code. By Lemma \ref{sig. code and B code}, there exist distinct codewords $\vec{c}_i,\vec{c}_j,\vec{c}_k,\vec{c}_s\in\C$ such that $\vec{c}_i+\vec{c}_j=\vec{c}_k+\vec{c}_s$. This implies that $\mathrm{supp}(\vec{c}_i)\cup \mathrm{supp}(\vec{c}_j)=\mathrm{supp}(\vec{c}_k)\cup\mathrm{supp}(\vec{c}_l)$, a contradiction to the assumption. Then the lemma follows. %that $\F$ is a weakly union-free family
\end{IEEEproof}

We remark that the converse of Lemma \ref{weakly and $B_2$} does not hold. That is, if $\C$ is a $2$-signature code, the corresponding family $\F$ may not be weakly union-free. The following is an example.

\begin{example}\label{not weakly UF}
Let $\C$ be a binary code of length $5$ and constant-weight $3$ defined as below.
\begin{equation*}
 \C=\begin{bmatrix}
1 & 0 & 1 & 0 \\
1 & 0 & 0 & 1 \\
1 & 1 & 1 & 0 \\
0 & 1 & 1 & 1 \\
0 & 1 & 0 & 1
\end{bmatrix}
\end{equation*}
Its corresponding family is $\F=\{\{1,2,3\},\{3,4,5\},\{1,3,4\},\{2,4,5\}\}$. It is easy to check that $\C$ is a $2$-signature code, but $\F$ is not weakly union-free since $\{1,2,3\}\cup\{3,4,5\}=\{1,3,4\}\cup\{2,4,5\}$.
\end{example}

Denote $F_3(n)$ as the maximum size of a weakly union-free family $\F\subseteq\binom{[n]}{3}$ and a weakly union-free family $\F\subseteq\binom{[n]}{3}$ is called \textit{optimal} if it has size $F_3(n)$. Then by Lemma \ref{weakly and $B_2$}, we have $A(n,3,2)\ge F_3(n)$. In \cite{C,CCL,FF}, the exact value of $F_3(n)=n(n-1)/3$ for any $n\ge 24$ and $n\in\{13,16,17,19,20,21\}$ was proved and direct constructions from combinatorial design theory for optimal weakly union-free families $\F\subseteq\binom{[n]}{3}$ were presented. Combining this with Theorem \ref{upper bound of $A(n,3,2)$} and Lemma \ref{weakly and $B_2$}, we have
%Frankl and F\"uredi \cite{FF} proved that for any $n\ge 3$, $F_3(n)\le n(n-1)/3$ and the equality holds for any sufficiently large integer $n$ with $n\equiv 1 \pmod 6$. Later,
%an upper bound for $F_3(n)$ and gave an explicit construction for optimal weakly union-free families by combinatorial design theory.

%\begin{lemma}\label{upper of $F_3(n)$}{\rm(\cite{CCL,C})}
%$F_3(n)\ge n(n-1)/3$ for any $n\ge 24$ and $n\in\{13,16,17,19,20,21\}$.
%\end{lemma}

%Therefore Lemma \ref{upper of $F_3(n)$}, together with Theorem \ref{upper bound of $A(n,3,2)$} and Lemma \ref{weakly and $B_2$}, implies

{\color{black}
\begin{theorem}\label{tight $A(n,3,2)$}
{\color{black}$A(n,3,2)=n(n-1)/3$ for any $n\ge 24$ and $n\in\{13,16,17,19,20,21\}$.}
\end{theorem}
}

We remark that by Theorem \ref{tight $A(n,3,2)$}, optimal $2$-signature codes of length $n$ and constant-weight $3$ for $n\ge 24$ and $n\in\{13,16,17,19,20,21\}$ can be constructed from optimal weakly union-free families $\F\subseteq\binom{[n]}{3}$. The interested reader is referred to \cite{C,CCL,FF} for more details on the constructions for optimal weakly union-free families $\F\subseteq\binom{[n]}{3}$.

\section{Constructions for $t$-signature codes {\color{black}with efficient decoding}}
\label{sec:4}

In this section, we provide two explicit constructions for signature codes, one is based on concatenation and the other on the Kronecker product. We prove that the signature codes obtained from these two constructions have efficient decoding algorithms and large code sizes. Moreover, we propose the concept of two-level signature code, and show that the product construction could be applied to construct two-level signature codes.

We first recap the definitions of frameproof code and superimposed code which will be exploited later. As stated in Section \ref{sec:1}, Cheng and Miao \cite{CM} proved that frameproof codes have an excellent traceability in a discretized model of multimedia fingerprinting, which promoted the study on frameproof codes, see \cite{SWGM} for example. Superimposed codes were proposed by Kautz and Singleton \cite{KS} for retrieving files, and later also extensively investigated in the contexts of \textit{disjunct matrices} and \textit{cover-free families}, see \cite{DH,EFF} for example.

Let $\C$ be an $(n,M,q)$ code. For any $1\le i\le n$, denote $\C(i)=\{c(i): \vec{c}=(c(1),\ldots,c(n))\in\C\}$ as the set of the $i$-th coordinates of $\C$. The {\it descendant code} of $\C$ is defined as
\begin{equation*}%\label{descendant}
  \mathrm{desc}(\C)=\C(1)\times\C(2)\times\cdots\times\C(n)\subseteq Q^n.
\end{equation*}

\begin{definition}\label{FPC}
Let $t\ge 2$ be an integer.
\begin{enumerate}
  \item $\C$ is an $(n,M,q)$ \textit{$t$-frameproof code}, or an $(n,M,q)$ $t$-FPC, if for any subset $\C_0\subseteq\C$ with $1\le|\C_0|\le t$, we have $\mathrm{desc}(\C_0)\cap\C=\C_0$, that is, for any other $\vec{c}=(c(1),\ldots,c(n))\in\C\setminus\C_0$, there exists $1\le k\le n$ such that $c(k)\not\in\C_0(k)$.

  \item $\C$ is an $(n,M,2)$ \textit{$t$-superimposed code}, if for any subset $\C_0\subseteq\C$ with $1\le|\C_0|\le t$ and any other codeword $\boldsymbol{c}=(c(1),\ldots,c(n))\in\C\setminus\C_0$, there exists $1\le k\le n$ such that $c(k)=1$ and $\C_0(k)=\{0\}$.
\end{enumerate}
\end{definition}

\subsection{Concatenated construction}
\label{subsec:4-1}

We now give the concatenated construction by taking a small $t$-signature code as the inner code
{\color{black}which guarantees the concatenated code to be a $t$-signature code, and taking a $q$-ary $t$-FPC as the outer code which makes sure the concatenated code can reduce complexity of decoding compared to an arbitrary $t$-signature code of the same length and size.
}

\begin{construction}\label{concatenated construction}
Let $\A$ be an $(n_1,M_1,q)$ $t$-FPC over the alphabet $Q$ and $\B$ be an $(n_2,q,2)$ $t$-signature code. Define a bijection $\phi:\ Q\rightarrow \B$. Let $\C$ be the code defined by
\begin{equation*}
\begin{split}
\Phi:\ \A & \rightarrow \C\\
\vec{a}=(a(1),\ldots,a(n_1)) & \mapsto\Phi(\vec{a})=(\phi(a(1)),\ldots,\phi(a(n_1))).
\end{split}
\end{equation*}
Then $\C$ is an $(n_1n_2,M_1,2)$ $t$-signature code.
\end{construction}
\begin{IEEEproof} It is obvious that $\C$ is an $(n_1n_2,M_1,2)$ code. For any distinct $\C_1,\C_2\subseteq\C$ with $1\le|\C_1|,|\C_2|\le t$, and for any real numbers $\lambda_j,\lambda_k'>0$ such that $\sum_{\vec{c}_j\in\C_1}\lambda_j=\sum_{\vec{c}_k\in\C_2}\lambda_k'=1$, we would like to show
\begin{equation}\label{concatenated sig. code}
\sum_{\vec{c}_j\in\C_1}\lambda_j\vec{c}_j\neq\sum_{\vec{c}_k\in\C_2}\lambda_k'\vec{c}_k.
\end{equation}
Since $\C_1\neq\C_2$, without loss of generality, we may assume that there exists one codeword $\boldsymbol{c}$ such that $\vec{c}\in\C_1\setminus\C_2$. Suppose that $\C_1,\C_2$ correspond to $\A_1,\A_2\subseteq\A$ respectively, and $\vec{c}$ corresponds to $\vec{a}=(a(1),\ldots,a(n_1))\in\A$. Clearly, $|\A_1|\le t$, $|\A_2|\le t$ and $\vec{a}\in\A_1\setminus\A_2$. Since $\A$ is an $(n_1,M_1,q)$ $t$-FPC, there must exist one coordinate $i$, $1\le i\le n_1$ such that $a(i)\not\in\A_2(i)$. This implies that $\A_1(i)\neq\A_2(i)$. Let {\color{black}$I_i=\{(i-1)n_2+1, (i-1)n_2+2,\ldots,in_2\}$}. Then
{\color{black}$\{\vec{c}_j|_{I_i}:\vec{c}_j\in\C_1\}$ and $\{\vec{c}_k|_{I_i}:\vec{c}_k\in\C_2\}$ are two multi-subsets of $\B$ with distinct base sets.} Since the inner code $\B$ is a $t$-signature code, by Proposition \ref{def-general-equiv} and the fact that $1\le|\C_1|,|\C_2|\le t$, we have
\begin{equation*}
\sum_{\vec{c}_j\in\C_1}\lambda_j\vec{c}_j|_{I_i}\neq\sum_{\vec{c}_k\in\C_2}\lambda_k'\vec{c}_k|_{I_i}
\end{equation*}
for any real numbers $\lambda_j,\lambda_k'>0$ such that $\sum_{\vec{c}_j\in\C_1}\lambda_j=\sum_{\vec{c}_k\in\C_2}\lambda_k'=1$, which implies (\ref{concatenated sig. code}). Hence $\C$ is a $t$-signature code.
\end{IEEEproof}

{\color{black}
In the literature, FPCs are usually constructed from structures in combinatorial design theory and coding theory such as \textit{orthogonal array}, \textit{packing designs} and \textit{error-correcting
codes}, see \cite{BT,B03,CZ,SW,SW2001} for example. As stated in Section \ref{sec2-3}, constructions for signature codes are only known in \cite{EFKL} where Goppa codes in coding theory were exploited. The following is an example to describe Construction \ref{concatenated construction} where we use a $2$-FPC constructed from an orthogonal array in \cite{BT} as the outer code and use a $2$-signature code of length $2$ shown in Corollary \ref{$A(2,2)$ and $A(3,2)$} as the inner code.

\begin{example}\label{Ex:concatenated sig}
Let $\A$ be a $(2,9,3)$ $2$-FPC and $\B$ be a $(2,3,2)$ $2$-signature code defined below.
\begin{equation}\label{Eq:ex-construction-1}
  \begin{matrix}
  \hspace{0.7cm} \vec{a}_1\hskip 0.15cm \vec{a}_2\hspace{0.15cm} \vec{a}_3\hspace{0.15cm} \vec{a}_4\hspace{0.15cm}\vec{a}_5\hspace{0.15cm} \vec{a}_6\hspace{0.15cm} \vec{a}_7\hspace{0.15cm} \vec{a}_8\hspace{0.15cm} \vec{a}_9\\
  \A=\begin{bmatrix}
   0 & 1 & 2 & 0 & 1 & 2 & 0 & 1 & 2 \\
   0 & 1 & 2 & 1 & 2 & 0 & 2 & 0 & 1 \\
   0 & 1 & 2 & 2 & 0 & 1 & 1 & 2 & 0
  \end{bmatrix},
  \end{matrix}\ \
  \begin{matrix}
  \hspace{0.7cm} \vec{b}_1\hskip 0.2cm \vec{b}_2\hspace{0.2cm} \vec{b}_3 \\
  \B=\begin{bmatrix}
   1 & 0 & 1  \\
   0 & 1 & 1
\end{bmatrix}.
\end{matrix}
\end{equation}
Define the bijection $\phi$ from $Q=\{0,1,2\}$ to $\B=\{\vec{b}_1,\vec{b}_2,\vec{b}_3\}$ as
\begin{equation}\label{Eq:bijection-1}
  \phi(k)=\vec{b}_{k+1},\ \forall k\in Q.
\end{equation}
By Construction \ref{concatenated construction}, the concatenated code $\C$ based on the outer code $\A$ and inner code $\B$ is a $(6,9,2)$ $2$-signature code presented below.
\begin{equation*}
  \begin{matrix}
  \hspace{0.55cm} \vec{c}_1\hskip 0.15cm \vec{c}_2\hspace{0.17cm} \vec{c}_3\hspace{0.17cm} \vec{c}_4\hspace{0.17cm}\vec{c}_5\hspace{0.17cm} \vec{c}_6\hspace{0.17cm} \vec{c}_7\hspace{0.19cm} \vec{c}_8\hspace{0.17cm} \vec{c}_9 \\
  \C=\begin{bmatrix}
   1 & 0 & 1 & 1 & 0 & 1 & 1 & 0 & 1 \\
   0 & 1 & 1 & 0 & 1 & 1 & 0 & 1 & 1 \\
   1 & 0 & 1 & 0 & 1 & 1 & 1 & 1 & 0 \\
   0 & 1 & 1 & 1 & 1 & 0 & 1 & 0 & 1 \\
   1 & 0 & 1 & 1 & 1 & 0 & 0 & 1 & 1 \\
   0 & 1 & 1 & 1 & 0 & 1 & 1 & 1 & 0
  \end{bmatrix}.
  \end{matrix}
\end{equation*}
\end{example}

We remark that the signature code obtained from Construction \ref{concatenated construction} has larger length and code size than the inner code which is also a signature code, but may be not good in terms of code rate. For example, the code rate of the concatenated signature code $\C$ in Example \ref{Ex:concatenated sig} is $(\log_2 9)/6\approx 0.5283$, while the code rate of the inner code $\B$ is $(\log_2 3)/2\approx 0.7925$. However, we will show that the concatenated signature code provides an efficient tracing algorithm which consists of two steps, that is, first decoding the inner code and then decoding the outer code.
}

\begin{theorem}\label{complexity 1}
The concatenated signature code $\C$ obtained by Construction \ref{concatenated construction} can trace back to a coalition of size at most $t$ in time $O(n_1n_2q^t+n_1M_1)$.
\end{theorem}
\begin{IEEEproof} Suppose that $\vec{r}=(\vec{r}_1,\vec{r}_2,\ldots,\vec{r}_{n_1})$ is an output of multimedia fingerprinting channel where $\vec{r}_i=(r_i(1),\ldots,r_i(n_2))$ for all $1\le i\le n_1$. Suppose that $\X\subseteq\C$, $|\X|\le t$ corresponds to the real coalition for the output $\vec{r}$. To determine $\X$ by $\vec{r}$ and the code $\C$, we divide the decoding process into the following two steps.

\textit{Step 1.} First we decode the inner code. For each $\vec{r}_i$, $1\le i\le n_1$, a subset $\B_i\subseteq\B$, $|\B_i|=s_i\le t$ can be traced back since the inner code $\B$ is a $t$-signature code. Denote $Q_i=\{\phi^{-1}(\vec{b}): \vec{b}\in\B_i\}\subseteq Q$, $1\le i\le n_1$. The time cost in this step is $O(n_1n_2q^t)$.

\textit{Step 2.} Next we decode the outer code. For the outer code $\A$, detect each codeword $\vec{a}=(a(1),\ldots,a(n_1))\in\A$ by checking if $a(i)\in Q_i$ for all $1\le i\le n_1$. If so, then the user corresponding to the codeword $\Phi(\vec{a})\in\C$ is identified as a colluder. Denote $\widehat{\X}=\{\Phi(\vec{a})\in\C: \vec{a}\in\A\ \text{and}\ a(i)\in Q_i, \forall 1\le i\le n_1\}$. The time cost in this step is $O(n_1M_1)$.

Now we show that $\X$ will be identified after Steps 1 and 2. To this end, we show that $\X=\widehat{\X}$. By Step 1, we have $\X\subseteq\widehat{\X}$. Assume that there exists $\vec{c}_0\in\widehat{\X}\setminus\X$. Then by Step 2, we have $\Phi^{-1}(\vec{c}_0)\in \mathrm{desc}(\{\Phi^{-1}(\vec{c}): \vec{c}\in\X\})$, a contradiction to the condition that $\A$ is a $t$-FPC. Hence we have $\X=\widehat{\X}$.

Based on Steps 1 and 2, the concatenated signature code $\C$ can trace back to a coalition of size at most $t$ in time $O(n_1n_2q^t+n_1M_1)$.
\end{IEEEproof}

{\color{black}
In general, as described in Section \ref{sec:signature code}, the decoding complexity of an $(n_1n_2,M_1,2)$ $t$-signature code is $O(n_1n_2M_1^t)$. According to Theorem \ref{complexity 1}, the $(n_1n_2,M_1,2)$ $t$-signature code obtained by Construction \ref{concatenated construction} can reduce the decoding complexity to $O(n_1n_2q^t+n_1M_1)$ if we choose an $(n_1,M_1,q)$ $t$-FPC with $q<M_1$ as the outer code. This can be achieved due to the known results on $t$-FPCs, see \cite{B03} for example. The following is an example to illustrate the decoding process described in Theorem \ref{complexity 1}.

\begin{example}
Consider the concatenated $(6,9,2)$ $2$-signature code $\C$ in Example \ref{Ex:concatenated sig} being applied to construct fingerprints for $9$ authorized users in multimedia fingerprinting. Each codeword $\vec{c}_j\in\C$ corresponds to the fingerprint of user $j$, $1\le j\le 9$. Suppose that $\X=\{\vec{c}_1,\vec{c}_9\}$ represents the coalition set and $\vec{r}$ is the output of multimedia fingerprinting channel generated by $\X$ as
\begin{equation*}
  \begin{matrix}
  \hspace{0.62cm} {\color{black}\vec{c}_1}\hskip 0.17cm \vec{c}_2\hspace{0.18cm} \vec{c}_3\hspace{0.18cm} \vec{c}_4\hspace{0.18cm}\vec{c}_5\hspace{0.18cm} \vec{c}_6\hspace{0.19cm} \vec{c}_7\hspace{0.19cm} \vec{c}_8\hspace{0.19cm} {\color{black}\vec{c}_9}\\
  \C=\begin{bmatrix}
   \begin{array}{ccccccccc}
   {\textit 1} & 0 & 1 & 1 & 0 & 1 & 1 & 0 & {\textit 1}  \\
   {\textit 0} & 1 & 1 & 0 & 1 & 1 & 0 & 1 & {\textit 1}  \\
   \hline%\hline
   {\textit 1} & 0 & 1 & 0 & 1 & 1 & 1 & 1 & {\textit 0}  \\
   {\textit 0} & 1 & 1 & 1 & 1 & 0 & 1 & 0 & {\textit 1}  \\
   \hline
   {\textit 1} & 0 & 1 & 1 & 1 & 0 & 0 & 1 & {\textit 1}  \\
   {\textit 0} & 1 & 1 & 1 & 0 & 1 & 1 & 1 & {\textit 0}
   \end{array}
  \end{bmatrix}
  \end{matrix}
  \begin{matrix}

                               \hspace{0.3cm} \vec{c}_1\hskip 0.35cm \vec{c}_2\hspace{0.35cm} \vec{c}_3\hspace{0.35cm} \vec{c}_4\hspace{0.35cm}\vec{c}_5\hspace{0.35cm} \vec{c}_6\hspace{0.35cm} \vec{c}_7\hspace{0.35cm} \vec{c}_8\hspace{0.35cm} \vec{c}_9\\
                               =\begin{bmatrix}
                               \begin{array}{ccccccccc}
                               {\color{black}\vec{b}_1} & \vec{b}_2 & {\color{black}\vec{b}_3} & {\color{black}\vec{b}_1} & \vec{b}_2 & {\color{black}\vec{b}_3} & {\color{black}\vec{b}_1} & \vec{b}_2 & {\color{black}\vec{b}_3} \\
                               \hline
                               {\color{black}\vec{b}_1} & {\color{black}\vec{b}_2} & \vec{b}_3 & {\color{black}\vec{b}_2} & \vec{b}_3 & {\color{black}\vec{b}_1} & \vec{b}_3 & {\color{black}\vec{b}_1} & {\color{black}\vec{b}_2}\\
                               \hline
                               {\color{black}\vec{b}_1} & \vec{b}_2 & \vec{b}_3 & \vec{b}_3 & {\color{black}\vec{b}_1} & \vec{b}_2 & \vec{b}_2 & \vec{b}_3 & {\color{black}\vec{b}_1}\\
                               \end{array}
                               \end{bmatrix}
                               \end{matrix}\ \underrightarrow{0.3\vec{c}_1+0.7\vec{c}_9}\
%  \begin{matrix}
%    \hspace{-0.4cm} \vec{r} \\
%    \begin{bmatrix}
%    \begin{array}{c}
%      1 \\
%      0.7 \\
%      \hline
%      0.3 \\
%      0.7 \\
%      \hline
%      1 \\
%      0
%      \end{array}
%    \end{bmatrix}=
%  \end{matrix}
  \begin{matrix}
  \hspace{-0.1cm} \vec{r}\\
  \begin{bmatrix}
  \begin{array}{c}
    \vec{r}_1 \\
    \hline
    \vec{r}_2 \\
    \hline
    \vec{r}_3
    \end{array}
  \end{bmatrix}
  \end{matrix}
\end{equation*}
where $\vec{r}_1=(1,0.7)^\top$, $\vec{r}_2=(0.3,0.7)^\top$ and $\vec{r}_3=(1,0)^\top$. We decode $\X$ by $\vec{r}$ and $\C$ in the following two steps.

\textit{Step 1.} As stated in Example \ref{Ex:concatenated sig}, the inner code $\B$ in (\ref{Eq:ex-construction-1}) is a $(2,3,2)$ $2$-signature code. For each $\vec{r}_i$, $1\le i\le 3$, we can decode that
\begin{equation*}
  \X|_{I_1}=\{\vec{b}_1,\vec{b}_3\},\ \X|_{I_2}=\{\vec{b}_1,\vec{b}_2\}\ \text{and}\ \X|_{I_3}=\{\vec{b}_1\}
\end{equation*}
where $I_1=\{1,2\}$, $I_2=\{3,4\}$ and $I_3=\{5,6\}$.

\textit{Step 2.} Denote $\Y\subseteq\A$ as the subset of the outer code corresponding to $\X$. Then by (\ref{Eq:ex-construction-1}) and (\ref{Eq:bijection-1}), we obtain
\begin{equation*}
  \Y(1)=\{0,2\},\ \Y(2)=\{0,1\}\ \text{and}\ \Y(3)=\{0\}.
\end{equation*}
To determine $\X$, it is equivalent to determine $\Y$. As stated in Example \ref{Ex:concatenated sig}, the outer code $\A$ is a $2$-FPC. We can decode $\Y=\{\vec{a}_1,\vec{a}_9\}$ by choosing each $\vec{a}_j$, $1\le j\le 9$, such that $a_j(i)\in\Y(i)$ for any $1\le i\le 3$.
\end{example}
}

\subsection{Product construction}
\label{subsec:4-2}

Now we provide a product construction by combining superimposed codes and signature codes as follows.
{\color{black}The ingredient $t$-signature code makes the constructed code $t$-signature, and the ingredient $t$-superimposed code allows for efficient decoding. Our constructed code can reduce decoding complexity compared to an arbitrary $t$-signature code of the same length and size.}

\begin{construction}\label{new construction}
Let $\A$ be an $(n_1,M_1,2)$ $t$-superimposed code, and $\B$ be an $(n_2,M_2,2)$ $t$-signature code and $\vec{0}\in\B$. Denote $\B^\ast=\B\setminus\{\vec{0}\}$. Let $\C=\A\otimes\B^\ast$, where $\A\otimes\B^\ast$ is the Kronecker product of $\A$ and $\B^\ast$:
\begin{equation}\label{product code}
\A\otimes\B^\ast=\left[
\begin{array}{cccc}
  {\color{black}a_1(1)}\B^\ast   & {\color{black}a_2(1)}\B^\ast   & \cdots & {\color{black}a_{\color{black}{M_1}}(1)}\B^\ast   \\
  {\color{black}a_1(2)}\B^\ast   & {\color{black}a_2(2)}\B^\ast   & \cdots & {\color{black}a_{\color{black}{M_1}}(2)}\B^\ast   \\
         \vdots                &      \vdots                  &        &         \vdots               \\
  {\color{black}a_1(n_1)}\B^\ast & {\color{black}a_2(n_1)}\B^\ast & \cdots & {\color{black}a_{\color{black}{M_1}}(n_1)}\B^\ast
\end{array}
\right].
\end{equation}
Then $\C$ is an $(n_1n_2,M_1(M_2-1),2)$ $t$-signature code.
\end{construction}
\begin{IEEEproof} It is obvious that $\B^\ast$ is an $(n_2,M_2-1,2)$ code, and thus $\C$ is an $(n_1n_2,M_1(M_2-1),2)$ code. For simplicity, we
{\color{black}divide $\C$ into $M_1$ groups of codewords as}
\begin{equation*}%\label{code group}
\C=\G_1\cup\G_2\cup\ldots\cup\G_{M_1}
\end{equation*}
where %for each group $\G_h$, $1\le h\le M_1$,
\begin{equation*}
  \G_h=
  \left[
  \begin{array}{c}
    a_h(1)\B^\ast \\
    a_h(2)\B^\ast \\
    \vdots \\
    a_h(n_1)\B^\ast
  \end{array}
  \right], \forall 1\le h\le M_1.
\end{equation*}

For any two distinct subsets $\C_1,\C_2\subseteq\C$ such that $|\C_1|\le t$ and $|\C_2|\le t$, denote $G_1=\{h\in[M_1]: \C_1\cap\G_h\ne\emptyset\}$ and $G_2=\{h\in[M_1]: \C_2\cap\G_h\ne\emptyset\}$. Obviously, $|G_1|\le t$ and $|G_2|\le t$. We would like to show that for any real numbers $\lambda_j,\lambda_k'>0$ such that $\sum_{\vec{c}_j\in\C_1}\lambda_j=\sum_{\vec{c}_k\in\C_2}\lambda_k'=1$, we have
\begin{equation}\label{product sig. code}
  \sum_{\vec{c}_j\in\C_1}\lambda_j\vec{c}_j\neq\sum_{\vec{c}_k\in\C_2}\lambda_k'\vec{c}_k.
\end{equation}
To this end, we consider the following two cases.
\begin{enumerate}
  \item $G_1=G_2$ and $|G_1|=|G_2|=s$, $1\le s\le t$.
  \begin{enumerate}
    \item[1.a)] If $s=1$, without loss of generality, we may assume $G_1=G_2=\{1\}$. Then $\C_1,\C_2\subseteq\G_1$. Since $\A$ is an $(n_1,M_1,2)$ $t$-superimposed code, $\A$ does not have $\vec{0}$ as a column vector. Then there must exist $i$, $1\le i\le n_1$ such that $a_1(i)=1$. Denote
        {\color{black}$I_i=\{(i-1)n_2+1, (i-1)n_2+2, \ldots, in_2\}$}, $1\le i\le n_1$. Recall that $\C_1\neq\C_2$. Then, according to (\ref{product code}), $\{\vec{c}_j|_{I_i}:\vec{c}_j\in\C_1\}$ and $\{\vec{c}_k|_{I_i}:\vec{c}_k\in\C_2\}$ are two distinct subsets of $\B^\ast$. Since $\B^\ast$ is a $t$-signature code, we have
       \begin{equation*}
         \sum_{\vec{c}_j\in\C_1}\lambda_j\vec{c}_j|_{I_i}\neq
         \sum_{\vec{c}_k\in\C_2}\lambda_k'\vec{c}_k|_{I_i}
       \end{equation*}
       for any real numbers $\lambda_j,\lambda_k'>0$ such that $\sum_{\vec{c}_j\in\C_1}\lambda_j=\sum_{\vec{c}_k\in\C_2}\lambda_k'=1$, implying (\ref{product sig. code}).

    \item[1.b)] If $2\le s\le t$, without loss of generality, we may assume $G_1=G_2=\{1,\ldots,s\}$. Then $|\C_1\cap\G_h|\le t-1$ and $|\C_2\cap\G_h|\le t-1$ for any $h\in G_1$. Since $\C_1\ne\C_2$, there exists at least one $h\in G_1$ such that $\C_1\cap\G_h\ne\C_2\cap\G_h$. Without loss of generality, we may assume $\C_1\cap\G_1\ne\C_2\cap\G_1$. Since $\A$ is a $t$-superimposed code, there exists $i$, $1\le i\le n_1$ such that $a_1(i)=1$ and $a_h(i)=0$ for any $2\le h\le s$. Clearly, $\vec{c}|_{I_i}=\vec{0}$ for any $\vec{c}\in\G_h$, $2\le h\le s$. Since $\B=\B^\ast\cup\{\vec{0}\}$ is a $t$-signature code, we have
        \begin{equation*}
          \sum_{\vec{c}_j\in\C_1\cap\G_1}\lambda_j\vec{c}_j|_{I_i}+\lambda\vec{0}\neq
          \sum_{\vec{c}_k\in\C_2\cap\G_1}\lambda_k'\vec{c}_k|_{I_i}+\lambda'\vec{0}
        \end{equation*}
        for any real numbers $\lambda_j,\lambda,\lambda_k',\lambda'>0$ such that $\sum_{\vec{c}_j\in\C_1\cap\G_1}\lambda_j+\lambda=\sum_{\vec{c}_k\in\C_2\cap\G_1}\lambda_k'+\lambda'=1$. This implies that
        \begin{equation*}
          \sum_{\vec{c}_j\in\C_1\cap\G_1}\lambda_j\vec{c}_j+\sum_{h=2}^s\sum_{\vec{c}_m\in\C_1\cap\G_h}\lambda_m\vec{c}_m\neq
          \sum_{\vec{c}_k\in\C_2\cap\G_1}\lambda_k'\vec{c}_k+\sum_{h=2}^s\sum_{\vec{c}_r\in\C_2\cap\G_h}\lambda_r'\vec{c}_r
        \end{equation*}
        for any real numbers $\lambda_j,\lambda_m,\lambda_k',\lambda_r'>0$ such that $\sum_{\vec{c}_j\in\C_1\cap\G_1}\lambda_j+\sum_{h=2}^s\sum_{\vec{c}_m\in\C_1\cap\G_h}\lambda_m=
        \sum_{\vec{c}_k\in\C_2\cap\G_1}\lambda_k'+\sum_{h=2}^s\sum_{\vec{c}_r\in\C_2\cap\G_h}\lambda_r'=1$, which further implies (\ref{product sig. code}).
  \end{enumerate}

  \item $G_1\ne G_2$. Without loss of generality, we may assume $G_1\setminus G_2\ne\emptyset$ and $h\in G_1\setminus G_2$. Since $\A$ is a $t$-superimposed code, there exists $i$, $1\le i\le n_1$ such that $a_h(i)=1$ and $a_{h'}(i)=0$ for all $h'\in G_2$. By the construction of $\C$ and the fact that $\B^\ast=\B\setminus\{\vec{0}\}$, $\C_1|_{I_i}$ contains at least one nonzero vector and $\C_2|_{I_i}=\{\boldsymbol{0}\}$, which implies that $\sum_{\vec{c}_j\in\C_1}\lambda_j\vec{c}_j|_{I_i}\neq\vec{0}$ and $\sum_{\vec{c}_k\in\C_2}\lambda_k'\vec{c}_k|_{I_i}=\vec{0}$ for any real numbers $\lambda_j,\lambda_k'>0$ such that $\sum_{\vec{c}_j\in\C_1}\lambda_j=
     \sum_{\vec{c}_k\in\C_2}\lambda_k'=1$. Then (\ref{product sig. code}) follows.
\end{enumerate}

Hence $\C$ is a $t$-signature code.
\end{IEEEproof}

{\color{black}
In the literature, constructions for superimposed codes were widely investigated. In \cite{KS}, Kautz and Singleton constructed superimposed codes from error-correcting codes such as \textit{maximum distance separable codes} in coding theory. Macula \cite{M1996} proposed a way of constructing disjunct matrices from a combinatorial viewpoint where the containment relation among sets was used. Combinatorial structures in combinatorial design theory such as packing designs and orthogonal arrays were used to construct cover-free families, see \cite{EFF,KS,SW} for example. In what follows, we use an example to illustrate Construction \ref{new construction} in which the $3\times 3$ identity matrix, which is a $(3,3,2)$ $2$-superimposed code by Definition \ref{FPC}, and the $(3,5,2)$ $2$-signature code shown in Example \ref{3-dim cube} are employed.

\begin{example}\label{Ex:product sig}
Let $\A$ be a $(3,3,2)$ $2$-superimposed code and $\B$ be a $(3,5,2)$ $2$-signature code defined below.
\begin{equation}\label{Eq:ex-construction-2}
  \begin{matrix}
  \hspace{0.7cm} \vec{a}_1\hskip 0.15cm \vec{a}_2\hspace{0.15cm} \vec{a}_3 \\
  \A=\begin{bmatrix}
   1 & 0 & 0  \\
   0 & 1 & 0  \\
   0 & 0 & 1
  \end{bmatrix},
  \end{matrix}\ \
  \begin{matrix}
  \hspace{0.7cm} \vec{b}_1\hskip 0.2cm \vec{b}_2\hspace{0.19cm} \vec{b}_3\hspace{0.18cm} \vec{b}_4\hspace{0.18cm} \vec{b}_5 \\
  \B=\begin{bmatrix}
   0 & 0 & 0 & 1 & 1  \\
   0 & 0 & 1 & 0 & 1  \\
   0 & 1 & 1 & 1 & 0
  \end{bmatrix}.
  \end{matrix}
\end{equation}
Then by Construction \ref{new construction}, $\C=\A\otimes\B^\ast$ is a $(9,12,2)$ $2$-signature code presented below.
\begin{equation*}
  \begin{matrix}
  \hspace{0.7cm} \vec{c}_1\hskip 0.2cm \vec{c}_2\hspace{0.2cm} \vec{c}_3\hspace{0.2cm} \vec{c}_4\hspace{0.2cm}\vec{c}_5\hspace{0.2cm} \vec{c}_6\hspace{0.2cm} \vec{c}_7\hspace{0.15cm} \vec{c}_8\hspace{0.1cm} \vec{c}_9 \hspace{0.1cm} \vec{c}_{10}\hspace{0.1cm} \vec{c}_{11}\hspace{0.1cm} \vec{c}_{12} \\
  \C=\begin{bmatrix}
  \begin{array}{cccc|cccc|cccc}
   0 & 0 & 1 & 1 & 0 & 0 & 0 & 0 & 0 & 0 & 0 & 0 \\
   0 & 1 & 0 & 1 & 0 & 0 & 0 & 0 & 0 & 0 & 0 & 0 \\
   1 & 1 & 1 & 0 & 0 & 0 & 0 & 0 & 0 & 0 & 0 & 0 \\
   \hline
   0 & 0 & 0 & 0 & 0 & 0 & 1 & 1 & 0 & 0 & 0 & 0 \\
   0 & 0 & 0 & 0 & 0 & 1 & 0 & 1 & 0 & 0 & 0 & 0 \\
   0 & 0 & 0 & 0 & 1 & 1 & 1 & 0 & 0 & 0 & 0 & 0 \\
   \hline
   0 & 0 & 0 & 0 & 0 & 0 & 0 & 0 & 0 & 0 & 1 & 1 \\
   0 & 0 & 0 & 0 & 0 & 0 & 0 & 0 & 0 & 1 & 0 & 1 \\
   0 & 0 & 0 & 0 & 0 & 0 & 0 & 0 & 1 & 1 & 1 & 0
  \end{array}
  \end{bmatrix}.
  \end{matrix}
\end{equation*}
\end{example}

We remark that the signature code $\C$ obtained from Construction \ref{new construction} has larger length and code size than the ingredient signature code $\B$, but may be not good in the sense of code rate. For example, the code rate of the signature code $\C$ in Example \ref{Ex:product sig} is $(\log_2 12)/9\approx 0.3983$, while the code rate of the ingredient signature code $\B$ is $(\log_2 5)/3\approx 0.7740$. However, we will show that the signature code obtained from Construction \ref{new construction} provides an efficient tracing algorithm which consists of two steps, that is, first decoding the ingredient superimposed code and then decoding the ingredient signature code. Notice that this is in some sense the converse order of the decoding in Theorem \ref{complexity 1}.
}

\begin{theorem}\label{complexity 2}
The signature code $\C$ obtained by Construction \ref{new construction} can trace back to a coalition of size at most $t$ in time  $O(tn_2M_2^t+n_1n_2M_1+tn_1)$.
\end{theorem}
\begin{IEEEproof} Suppose that $\vec{r}=(\vec{r}_1,\ldots,\vec{r}_{n_1})$ is an output of the multimedia fingerprinting channel where $\vec{r}_i=(r_i(1),\ldots,r_i(n_2))$ for all $1\le i\le n_1$. Suppose that $\X\subseteq\C$, $|\X|\le t$ corresponds to the real coalition for the output $\vec{r}$. Denote $G=\{h\in [M_1]: \X\cap\G_h\ne\emptyset\}$ as the set of group indices of the codewords in $\X$. Obviously, $|G|\le t$. To determine $\X$ by $\vec{r}$ and the code $\C$, we will first determine $G$, and then determine $\X\cap\G_h$ for any $h\in G$.

\textit{Step 1.} For the ingredient superimposed code $\A$, detect each codeword $(a_h(1),\ldots,a_h(n_1))\in\A$, $1\le h\le M_1$ by checking if there exists $i$, $1\le i\le n_1$ such that $\vec{r}_i=\vec{0}$ and $a_h(i)=1$. Denote $G_0=\{h\in[M_1]: \exists 1\le i\le n_1\ \text{s.t.}\ \vec{r}_i=\vec{0}\ \text{and}\ a_h(i)=1\}$ and $\widehat{G}=[M_1]\setminus G_0$. We claim that $G=\widehat{G}$. The time cost in this step is $O(n_1n_2M_1)$.

To verify our claim, we first show $G\subseteq\widehat{G}$. Assume that there exists $h\in G$ but $h\not\in\widehat{G}$, that is, $h\in G_0$. Then there exists $i$, $1\le i\le n_1$ such that $\vec{r}_i=\vec{0}$ and $a_h(i)=1$. Notice that $\vec{r}_i=\vec{0}$ implies $\X|_{I_i}=\{\vec{0}\}$. Then we have $\X\cap\G_h=\emptyset$, that is, $h\not\in G$, a contradiction to the assumption. So we have $G\subseteq\widehat{G}$. On the other hand, we show $\widehat{G}\subseteq G$. Assume that there exists $h'\in\widehat{G}$ but $h'\not\in G$. Since $\A$ is a $t$-superimposed code, there exists $i'$, $1\le i'\le n_1$ such that $a_{h'}(i')=1$ and $a_h(i')=0$ for any $h\in G$. Recall that $G$ is the set of group indices of the codewords in $\X$. Thus we have $\X|_{I_{i'}}=\{\vec{0}\}$ which implies $\vec{r}_{i'}=\vec{0}$. Then we have $h'\in G_0$, a contradiction to the assumption that $h'\in\widehat{G}$ since $\widehat{G}\cap G_0=\emptyset$. So, we have $G=\widehat{G}$ and thus $G$ can be determined after Step 1.

\textit{Step 2.} In this step, we show how to determine $\X\cap\G_h$ for any $h\in G$.
\begin{enumerate}
  \item If $|G|=1$, without loss of generality, we may assume $G=\{1\}$. Then $\X=\X\cap\G_1$. Since $\A$ is a $t$-superimposed code, there exists $i$, $1\le i\le n_1$ such that $a_1(i)=1$. Then we have $\vec{r}_{i}\ne\vec{0}$. Since $\B^{\ast}$ is a $t$-signature code, we can determine $\X|_{I_i}$ by $\vec{r}_{i}$, and thus can determine $\X$. The time cost in this case is $O(n_2(M_2-1)^t+n_1)$.

  \item If $|G|=s\ge 2$, then we have $2\le\sum_{h\in G}|\X\cap\G_h|\le t$, and $1\le|\X\cap\G_h|\le t-1$ for any $h\in G$. Since $\A$ is a $t$-superimposed code, for any $h\in G$, there exists $i$, $1\le i\le n_1$ such that $a_h(i)=1$ and $a_{h'}(i)=0$ for any $h'\in G\setminus\{h\}$. Then we have $(\X\setminus\G_h)|_{I_i}=\{\vec{0}\}$. Since $\B=\B^{\ast}\cup\{\vec{0}\}$ is a $t$-signature code, we can determine $(\X\cap\G_h)|_{I_i}$ by $\vec{r}_{i}$, and thus can determine $\X\cap\G_h$ for any $h\in G$. The time cost in this case is $O(tn_2M_2^t+tn_1)$.
\end{enumerate}

Based on Steps 1 and 2, the $t$-signature code obtained by Construction \ref{new construction} can trace back to all the colluders of size no more than $t$ in time $O(tn_2M_2^t+n_1n_2M_1+tn_1)$.
\end{IEEEproof}

In general, the decoding complexity of an $(n_1n_2,M_1(M_2-1),2)$ $t$-signature code is $O(n_1n_2M_1^t(M_2-1)^t)$. Here according to Theorem \ref{complexity 2}, we can reduce the decoding complexity to $O(tn_2M_2^t+n_1n_2M_1+tn_1)$ by using the $t$-signature code obtained from Construction \ref{new construction}. The following is an example to show the decoding process described in Theorem \ref{complexity 2}.

{\color{black}
\begin{example}\label{Ex:product sig-2}
Consider the $(9,12,2)$ $2$-signature code $\C$ in Example \ref{Ex:product sig} being applied to construct fingerprints for $12$ authorized users in multimedia fingerprinting. Each codeword $\vec{c}_j\in\C$ corresponds to the fingerprint of user $j$, $1\le j\le 12$. We divide all the codewords of $\C$ into $3$ groups $\C=\G_1\cup\G_2\cup\G_3$ where $\G_1=\{\vec{c}_1,\ldots,\vec{c}_4\}$, $\G_2=\{\vec{c}_5,\ldots,\vec{c}_8\}$ and $\G_3=\{\vec{c}_9,\ldots,\vec{c}_{12}\}$, and divide all the coordinates of the codewords in $\C$ into $ I_1=\{1,2,3\}$, $I_2=\{4,5,6\}$ and $I_3=\{7,8,9\}$. Suppose that $\X\subseteq\C$ with $|\X|\le 2$ represents the coalition set, $\vec{r}$ is the output of multimedia fingerprinting channel generated by $\X$, and $G=\{1\le h\le 3: \X\cap \G_h\neq\emptyset\}$ is the set of group indices of the codewords in $\X$. We discuss two cases on the distributions of codewords in $\X$.
\begin{enumerate}%[{\rm 1)}]
  \item[1)] Suppose that $\X=\{\vec{c}_2,\vec{c}_4\}$ and the output $\vec{r}$ generated by $\X$ is
  \begin{equation*}
  \begin{matrix}
  \hspace{0.75cm} \vec{c}_1\hskip 0.2cm {\color{black}\vec{c}_2}\hspace{0.2cm} \vec{c}_3\hspace{0.2cm} {\color{black}\vec{c}_4}\hspace{0.2cm}\vec{c}_5\hspace{0.2cm} \vec{c}_6\hspace{0.2cm} \vec{c}_7\hspace{0.15cm} \vec{c}_8\hspace{0.15cm} \vec{c}_9 \hspace{0.1cm} \vec{c}_{10}\hspace{0.1cm} \vec{c}_{11}\hspace{0.1cm} \vec{c}_{12} \\
  \C=\begin{bmatrix}
  \begin{array}{cccc|cccc|cccc}
   0 & {\textit 0} & 1 & {\textit 1} & 0 & 0 & 0 & 0 & 0 & 0 & 0 & 0 \\
   0 & {\textit 1} & 0 & {\textit 1} & 0 & 0 & 0 & 0 & 0 & 0 & 0 & 0 \\
   1 & {\textit 1} & 1 & {\textit 0} & 0 & 0 & 0 & 0 & 0 & 0 & 0 & 0 \\
   \hline
   0 & {\textit 0} & 0 & {\textit 0} & 0 & 0 & 1 & 1 & 0 & 0 & 0 & 0 \\
   0 & {\textit 0} & 0 & {\textit 0} & 0 & 1 & 0 & 1 & 0 & 0 & 0 & 0 \\
   0 & {\textit 0} & 0 & {\textit 0} & 1 & 1 & 1 & 0 & 0 & 0 & 0 & 0 \\
   \hline
   0 & {\textit 0} & 0 & {\textit 0} & 0 & 0 & 0 & 0 & 0 & 0 & 1 & 1 \\
   0 & {\textit 0} & 0 & {\textit 0} & 0 & 0 & 0 & 0 & 0 & 1 & 0 & 1 \\
   0 & {\textit 0} & 0 & {\textit 0} & 0 & 0 & 0 & 0 & 1 & 1 & 1 & 0
  \end{array}
  \end{bmatrix}
  \end{matrix}\ \underrightarrow{0.4\vec{c}_2+0.6\vec{c}_4}\
%  \begin{matrix}
%  \hspace{-0.4cm} \vec{r} \\
%  \begin{bmatrix}
%  \begin{array}{c}
%      0.6 \\
%      1   \\
%      0.4 \\
%      \hline
%      0 \\
%      0 \\
%      0 \\
%      \hline
%      0 \\
%      0 \\
%      0
%      \end{array}
%    \end{bmatrix}\triangleq
%  \end{matrix}
  \begin{matrix}
   \hspace{-0.1cm} \vec{r} \\
  \begin{bmatrix}
  \begin{array}{c}
    \vec{r}_1 \\
    \hline
    \vec{r}_2 \\
    \hline
    \vec{r}_3
    \end{array}
  \end{bmatrix}
  \end{matrix}
  \end{equation*}
  where $\vec{r}_1=(0.6,1,0.4)^\top$ and $\vec{r}_2=\vec{r}_3=(0,0,0)^\top$. We decode $\X$ by $\vec{r}$ and $\C$ in the following two steps.

  \textit{Step 1.} As shown in (\ref{Eq:ex-construction-2}) in Example \ref{Ex:product sig}, $\A$ is a $2$-superimposed code and $a_2(2)=a_3(3)=1$. Since  $\vec{r}_2=\vec{r}_3=(0,0,0)^\top$, we can decode $G=\{1\}$, implying that $\X\subseteq\G_1$.

  \textit{Step 2.} Based on Step 1, we focus on $\G_1$. As shown in (\ref{Eq:ex-construction-2}), $\B^\ast$ is a $2$-signature code, then we can decode $\X|_{I_1}=\{\vec{b}_3,\vec{b}_5\}$ by $\vec{r}_1$, which implies $\X=\{\vec{c}_2,\vec{c}_4\}$.

  \item[{\rm 2)}] Suppose that $\X=\{\vec{c}_2,\vec{c}_7\}$ and the output $\vec{r}$ generated by $\X$ is
  \begin{equation*}
  \begin{matrix}
  \hspace{0.78cm} \vec{c}_1\hskip 0.18cm {\color{black}\vec{c}_2}\hspace{0.2cm} \vec{c}_3\hspace{0.2cm} \vec{c}_4\hspace{0.2cm}\vec{c}_5\hspace{0.2cm} \vec{c}_6\hspace{0.2cm} {\color{black}\vec{c}_7}\hspace{0.15cm} \vec{c}_8\hspace{0.18cm} \vec{c}_9 \hspace{0.12cm} \vec{c}_{10}\hspace{0.12cm} \vec{c}_{11}\hspace{0.1cm} \vec{c}_{12} \\
  \C=\begin{bmatrix}
  \begin{array}{cccc|cccc|cccc}
   0 & {\textit 0} & 1 & 1 & 0 & 0 & {\textit 0} & 0 & 0 & 0 & 0 & 0 \\
   0 & {\textit 1} & 0 & 1 & 0 & 0 & {\textit 0} & 0 & 0 & 0 & 0 & 0 \\
   1 & {\textit 1} & 1 & 0 & 0 & 0 & {\textit 0} & 0 & 0 & 0 & 0 & 0 \\
   \hline
   0 & {\textit 0} & 0 & 0 & 0 & 0 & {\textit 1} & 1 & 0 & 0 & 0 & 0 \\
   0 & {\textit 0} & 0 & 0 & 0 & 1 & {\textit 0} & 1 & 0 & 0 & 0 & 0 \\
   0 & {\textit 0} & 0 & 0 & 1 & 1 & {\textit 1} & 0 & 0 & 0 & 0 & 0 \\
   \hline
   0 & {\textit 0} & 0 & 0 & 0 & 0 & {\textit 0} & 0 & 0 & 0 & 1 & 1 \\
   0 & {\textit 0} & 0 & 0 & 0 & 0 & {\textit 0} & 0 & 0 & 1 & 0 & 1 \\
   0 & {\textit 0} & 0 & 0 & 0 & 0 & {\textit 0} & 0 & 1 & 1 & 1 & 0
  \end{array}
  \end{bmatrix}
  \end{matrix}\ \underrightarrow{0.5\vec{c}_2+0.5\vec{c}_7}\
%  \begin{matrix}
%  \hspace{-0.4cm} \vec{r} \\
%  \begin{bmatrix}
%  \begin{array}{c}
%      0 \\
%      0.5   \\
%      0.5 \\
%      \hline
%      0.5\\
%      0 \\
%      0.5 \\
%      \hline
%      0 \\
%      0 \\
%      0
%      \end{array}
%    \end{bmatrix}\triangleq
%  \end{matrix}
  \begin{matrix}
   \hspace{-0.1cm} \vec{r} \\
  \begin{bmatrix}
  \begin{array}{c}
    \vec{r}_1 \\
    \hline
    \vec{r}_2 \\
    \hline
    \vec{r}_3
    \end{array}
  \end{bmatrix}
  \end{matrix}
  \end{equation*}
  where $\vec{r}_1=(0,0.5,0.5)^\top$, $\vec{r}_2=(0.5,0,0.5)^\top$ and $\vec{r}_3=(0,0,0)^\top$. We decode $\X$ by $\vec{r}$ and $\C$ in the following two steps.

  \textit{Step 1.} By $\vec{r}_3=(0,0,0)^\top$ and $a_3(3)=1$, we first can decode $G=\{1,2\}$, implying that $\X=(\X\cap\G_1)\cup(\X\cap\G_2)$.

  \textit{Step 2.} In this step, we determine $\X\cap\G_1$ and $\X\cap\G_2$. Note that $\B=\B^\ast\cup\{\vec{0}\}$ is a $2$-signature code. Then by $\vec{r}_1$, we can determine $\X|_{I_1}=\{\vec{0},\vec{b}_3\}$. Since $a_2(1)=0$, we have $(\X\cap\G_2)|_{I_1}=\{\vec{0}\}$, implying $(\X\cap\G_1)|_{I_1}=\{\vec{b}_3\}$. From the construction of $\C$ in Example \ref{Ex:product sig}, we obtain $\X\cap\G_1=\{\vec{c}_2\}$. Similarly, we can decode $\X\cap\G_2=\{\vec{c}_7\}$ by $\vec{r}_2$. Thus, we have $\X=\{\vec{c}_2,\vec{c}_7\}$.
\end{enumerate}
\end{example}

}

\subsection{Two-level signature code}
\label{subsec:4-3}

In this subsection, we show that Construction \ref{new construction} can be applied to construct two-level signature codes as well.
{\color{black}Two-level fingerprinting codes were first investigated by Anthapadmanabhan and Barg \cite{AB2009} in digital fingerprinting for the purpose of getting partial information if the size of a coalition exceeds a predetermined threshold. As shown in \cite{AB2009}, users accommodated in the digital fingerprinting system were partitioned into several groups and \textit{two-level $(t_1,t_2)$-traceability codes} with $t_1>t_2$ were introduced to guarantee that once a pirate copy produced by a coalition is confiscated, the dealer can: 1) identify at least one colluder if the coalition size is no more than $t_2$, which is the same to a traditional $t_2$-\textit{traceability code} (one-level), and 2) trace back to at least one group that contains at least one colluder if the coalition size is larger than $t_2$ but no more than $t_1$. Inspired by the idea and the applications of two-level traceability codes, in the literature, several other types of two-level fingerprinting codes were proposed, see \cite{AB2010,R2011} for example. Here, we introduce the concept of \textit{two-level signature code} for multimedia fingerprinting. To the best of our knowledge, this is the first work to introduce this concept.
}

%Two-level codes were introduced by Anthapadmanabhan and Barg \cite{AB2009} in digital fingerprinting with the feature that all the codewords are classified into groups and each group consists of several codewords. They showed that a two-level $(t_1,t_2)$-fingerprinting code, where $t_1>t_2$, could identify all the colluders in digital fingerprinting if the coalition size is no more than $t_2$, and determine all the groups containing at least one colluder if the coalition size is no more than $t_1$. Inspired by this, we introduce the concept of two-level signature codes for multimedia fingerprinting.

\begin{definition}\label{two-level}
Let $\C$ be an $(n,M,2)$ code and $t_1,t_2$ be positive integers with $t_1>t_2$. Suppose that all the codewords of $\C$ are partitioned into $M_1(<M)$ groups: $\C=\G_1\cup\G_2\cup\ldots\cup\G_{M_1}$. For any subset $\C'\subseteq\C$, denote $G(\C')=\{h\in [M_1]: \C'\cap\G_h\ne\emptyset\}$ as the set of group indices of the codewords in $\C'$. $\C$ is an $(n,M,2)$ \textit{two-level $(t_1,t_2)$-signature code} if for any two subsets $\C_1,\C_2\subseteq\C$ and for any real numbers $\lambda_j,\lambda_k'>0$ such that $\sum_{\vec{c}_j\in\C_1}\lambda_j=\sum_{\vec{c}_k\in\C_2}\lambda_k'=1$,
\begin{equation*}
  \sum_{\vec{c}_j\in\C_1}\lambda_j\vec{c}_j=\sum_{\vec{c}_k\in\C_2}\lambda_k'\vec{c}_k
\end{equation*}
implies that
\begin{enumerate}
  \item $\C_1=\C_2$ if $1\le |\C_1|,|\C_2|\le t_2$;
  \item $G(\C_1)=G(\C_2)$ if $1\le |\C_1|,|\C_2|\le t_1$.
\end{enumerate}
\end{definition}

By Definition \ref{two-level}, in multimedia fingerprinting, a two-level $(t_1,t_2)$-signature code could
\begin{enumerate}
  \item[a)] trace back to all the colluders if the coalition size is at most $t_2$;
  \item[b)] determine all the groups each of which contains at least one colluder if the coalition size is at most $t_1$.
\end{enumerate}

We have the following relationship between signature codes (\textit{one-level}) and two-level signature codes.

{\color{black}
\begin{lemma}\label{Lem:two-level}
Let $t_1,t_2$ be two positive integers such that $t_1>t_2$. Then a {\color{black}$t_1$-signature code is a two-level $(t_1,t_2)$-signature code, and a two-level $(t_1,t_2)$-signature code is a $t_2$-signature code.}
\end{lemma}
{\color{black}
\begin{IEEEproof} By Definitions \ref{sig. code} and \ref{two-level}, it is obvious that a two-level $(t_1,t_2)$-signature code is a $t_2$-signature code.

Suppose that $\C$ is a $t_1$-signature code, then we can partition all the codewords of $\C$ into non-empty groups.
\begin{enumerate}
  \item By Definition \ref{sig. code} and the condition that $t_1>t_2$, $\C$ is a $t_2$-signature code;

  \item For any two subsets $\C_1,\C_2\subseteq\C$, if $1\le |\C_1|,|\C_2|\le t_1$ and the group indices of $\C_1,\C_2$ are different, then $\C_1\neq\C_2$. By Definition
  \ref{sig. code}, for any real numbers $\lambda_j,\lambda_k'>0$ such that $\sum_{\vec{c}_j\in\C_1}\lambda_j=\sum_{\vec{c}_k\in\C_2}\lambda_k'=1$, we have $\sum_{\vec{c}_j\in\C_1}\lambda_j\vec{c}_j\neq\sum_{\vec{c}_k\in\C_2}\lambda_k'\vec{c}_k$.
\end{enumerate}
Thus, by Definition \ref{two-level}, $\C$ is a $(t_1,t_2)$-signature code. The lemma follows.
\end{IEEEproof}
}
}

Lemma \ref{Lem:two-level} implies that a $(t_1,t_2)$-signature code lies between a {\color{black}$t_1$}-signature code and a $t_2$-signature code.
The following construction for two-level signature codes is an application of Construction \ref{new construction}.

\begin{theorem}
Let $t_1,t_2$ be two positive integers such that $t_1>t_2$. Let $\A$ be an $(n_1,M_1,2)$ $t_1$-superimposed code and $\B$ be an $(n_2,M_2,2)$ $t_2$-signature code containing $\vec{0}$. Let $\B^\ast=\B\setminus\{\vec{0}\}$. Then {\color{black}$\C =\A\otimes\B^\ast$ defined by (\ref{product code}) is} an $(n_1n_2,M_1(M_2-1),2)$ two-level $(t_1,t_2)$-signature code with $M_1$ groups.
\end{theorem}
\begin{IEEEproof} By Construction \ref{new construction}, an $(n_1n_2,M_1(M_2-1),2)$ code $\C$ can be obtained from $\A$ and $\B$. Now we show that $\C$ is a two-level $(t_1,t_2)$-signature code. It is obvious that all the codewords of $\C$ are divided into $M_1$ groups and each group contains $M_2-1$ codewords. If a forged copy is created by a coalition of size at most $t_2$, all the colluders will be identified by Theorem \ref{complexity 2}. If a forged copy is created by a coalition of size at most $t_1$, by Step 1 of the decoding process in the argument of Theorem \ref{complexity 2}, any group containing at least one colluder will be identified. The proof is completed.
\end{IEEEproof}

\section{Comments on noisy weighted binary adder channel and multimedia fingerprinting}
\label{sec:5}

In the previous sections, we investigated signature codes for noiseless multimedia fingerprinting channel. In this section, we consider the noisy scenario. First we show from a geometric viewpoint that there does not exist any binary code with complete traceability for noisy multimedia fingerprinting channel. Then we introduce frameproof signature codes for noisy multimedia fingerprinting to protect innocent {\color{black}groups} of users by guaranteeing that disjoint coalition sets cannot produce an identical forged copy.

\subsection{Noisy multimedia fingerprinting}
\label{subsec:5-2}

In practice, the noisy multimedia fingerprinting channel is more realistic but with more {\color{black}complicated} assumptions than the noiseless case. That is, the dealer observes the forged copy with some noise which may be produced artificially by the coalition before redistributing the forged copy, or generated unartificially during the redistribution process of the forged copy. We show from a geometric viewpoint that no coalition could be completely traced back in noisy multimedia fingerprinting.%there does not exist any binary code with complete traceability for noisy multimedia fingerprinting channel, that is,  %Signature codes were considered for noiseless multimedia fingerprinting channel, and an equivalent geometric description of signature codes was shown in Proposition \ref{noiseless geometric}. However,

Assume that the colluders in a coalition $J$ add some noise $\vec{\varepsilon}$ for the purpose of making themselves less likely to be identified, where $\vec{\varepsilon}=(\varepsilon(1),\ldots,\varepsilon(m))\in\mathbb{R}^m\setminus\{\vec{0}\}$. {\color{black}Typically, we consider the adversarial noise with bounded energy, that is, $\|\vec{\varepsilon}\|< \delta$ for some real number $\delta>0$.}
%such that $\|\vec{\varepsilon}\|<\delta$ and $\delta$ is very small.
Then the dealer observes the forged copy
\begin{equation}\label{noisy forged copy}
\tilde{\vec{y}}=\vec{x}+\sum_{j\in J}\lambda_j\vec{w}_j+\vec{\varepsilon}.
\end{equation}
Notice that $\vec{\varepsilon}$ is chosen by the coalition $J$ and unknown to the dealer. However, the dealer can calculate
\begin{equation}\label{noisy $r_k$}
  \tilde{r}(k)=\langle\tilde{\vec{y}}-\vec{x},\vec{f}_k\rangle
\end{equation}
for $1\le k\le n$, and obtain
\begin{equation}\label{noisy $r$}
\tilde{\vec{r}}=(\tilde{r}(1),\ldots,\tilde{r}(n))=\sum\limits_{j\in J}\lambda_j\vec{c}_j+\vec{e}
\end{equation}
where $\vec{e}=(e(1),\ldots,e(n))\in\mathbb{R}^n$ and $e(k)=\langle\vec{\varepsilon},\vec{f}_k\rangle$ for $1\le k\le n$. It should be noted that the dealer does not know what $\vec{e}$ is, but knows that $\|\vec{e}\|\le\|\vec{\varepsilon}\|<\delta$.

%%%%%%%%%%%%%%%%%%%%%%%%%%%%%%%%%%%%%%%%%%%%%%%%%%%%%%%%%%%%%%%%%%%%%%%%%%%%%%%%%%%%%%%%
%{\color{brown}(You have already said $\|\vec{e}\|< \delta$ before. Should you say  $\|\vec{e}\|< \delta$ here? (not $\|\vec{e}\|\le \|\vec{\varepsilon}\|$?!))}

%It seems that we never said $\|\vec{e}\|< \delta$ before.

% I think if bound of noise energy $\delta$ is publicly well-known (here, noise is chosen by the coalition), then we can say that `the dealer knows $\|\vec{e}\|\le\|\vec{\varepsilon}\|<\delta$'. Otherwise, we can only say `the dealer knows $\|\vec{e}\|\le\|\vec{\varepsilon}\|'. However, before designing a code for noisy MF, the dealer can always assume that $\|\vec{\varepsilon}\|<\delta$ (or $<\gamma$ etc.). Combining with the knowledge, i.e., $\|\vec{e}\|\le\|\vec{\varepsilon}\|, the dealer can define codes under the condition $\|\vec{e}\|<\delta$.
%%%%%%%%%%%%%%%%%%%%%%%%%%%%%%%%%%%%%%%%%%%%%%%%%%%%%%%%%%%%%%%%%%%%%%%%%%%%%%%%%%%%%%%%

In noisy multimedia fingerprinting, the dealer would also like to design a binary code $\C$ with some properties to identify the whole coalition set $J$ based on the result $\tilde{\vec{r}}$ calculated in (\ref{noisy $r$}). We define the complete traceability of a binary code for noisy multimedia fingerprinting as follows.

{\color{black}
\begin{definition}\label{complete traceability}
\color{black}
Let $\C$ be an $(n,M,2)$ code, $t\ge 2$ be an integer and $\delta>0$ be a real number. $\C$ has \textit{$(t,\delta)$-complete traceability} if
for any two distinct subsets $\C_1,\C_2\subseteq\C$ with $1\le|\C_1|,|\C_2|\le t$, we have
\begin{equation*}\label{noisy sig. code}
\sum_{\vec{c}_j\in\C_1}\lambda_j\vec{c}_j+\vec{e}\neq\sum_{\vec{c}_k\in\C_2}\lambda_k'\vec{c}_k+\vec{e}'
\end{equation*}
for any $\vec{e},\vec{e}'\in\mathbb{R}^n$ and any real numbers $\lambda_j,\lambda_k'>0$ such that $\|\vec{e}\|,\|\vec{e}'\|<\delta$ and $\sum_{\vec{c}_j\in\C_1}\lambda_j=\sum_{\vec{c}_k\in\C_2}\lambda_k'=1$.
\end{definition}
}

Let $\vec{z}\in\mathbb{R}^n$ be a point in the $n$-dimensional Euclidean space and $\delta>0$ be a real number. An \textit{open $n$-ball with center $\vec{z}$ and radius $\delta$} is formed by all the points in $\{\vec{z}'\in\mathbb{R}^n: \|\vec{z}-\vec{z}'\|<\delta\}$.
For any distinct $S_1,S_2\subseteq\mathbb{R}^n$, define the \textit{distance} between $S_1$ and $S_2$ as
\begin{equation}\label{general distance}
  \mathrm{d}(S_1,S_2)=\inf\{\|\vec{z}-\vec{z}'\|: \vec{z}\in S_1, \vec{z}'\in S_2\}.
\end{equation}
Note that for any distinct $S_1,S_2\subseteq\mathbb{R}^n$, if $S_1\cap S_2\neq\emptyset$, then we must have $\mathrm{d}(S_1,S_2)=0$. We have the following equivalent description for Definition \ref{complete traceability}.

% However, if $S_1\cap S_2=\emptyset$, it is also possible that $\mathrm{d}(S_1,S_2)=0$.

\begin{proposition}\label{noisy geometric}
An $(n,M,2)$ code $\C$ has $(t,\delta)$-complete traceability if and only if for any two distinct subsets $\C_1,\C_2\subseteq\C$ with $1\le|\C_1|,|\C_2|\le t$, we have
\begin{equation}\label{distance}
\mathrm{d}(\mathcal{P}(\C_1),\mathcal{P}(\C_2))\ge 2\delta.
\end{equation}
\end{proposition}
\begin{IEEEproof}
{\color{black}
Let $\C$ be an $(n,M,2)$ code, $\C_1,\C_2\subseteq\C$ be two distinct subsets with $1\le|\C_1|,|\C_2|\le t$ and $\lambda_j,\lambda_k'>0$ be real numbers such that $\sum_{\vec{c}_j\in\C_1}\lambda_j=\sum_{\vec{c}_k\in\C_2}\lambda_k'=1$. Then $\sum_{\vec{c}_j\in\C_1}\lambda_j\vec{c}_j$ is a point in $\mathcal{P}(\C_1)$ and $\sum_{\vec{c}_k\in\C_2}\lambda_k'\vec{c}_k$ is a point in $\mathcal{P}(\C_2)$. Moreover, for any $\vec{e},\vec{e}'\in\mathbb{R}^n$ with $\|\vec{e}\|,\|\vec{e}'\|<\delta$, $\sum_{\vec{c}_j\in\C_1}\lambda_j\vec{c}_j+\vec{e}$ is a point in the open $n$-ball with center $\sum_{\vec{c}_j\in\C_1}\lambda_j\vec{c}_j$ and radius $\delta$, and $\sum_{\vec{c}_k\in\C_2}\lambda_k'\vec{c}_k+\vec{e}'$ is a point in the open $n$-ball with center $\sum_{\vec{c}_k\in\C_2}\lambda_k'\vec{c}_k$ and radius $\delta$. Then the argument
\begin{equation*}
  \sum_{\vec{c}_j\in\C_1}\lambda_j\vec{c}_j+\vec{e}\neq\sum_{\vec{c}_k\in\C_2}\lambda_k'\vec{c}_k+\vec{e}',\ \forall \vec{e},\vec{e}'\in\mathbb{R}^n\ \text{with}\ \|\vec{e}\|,\|\vec{e}'\|<\delta
\end{equation*}
holds if and only if
\begin{equation}\label{distance-1}
  \Big\|\sum_{\vec{c}_j\in\C_1}\lambda_j\vec{c}_j-\sum_{\vec{c}_k\in\C_2}\lambda_k'\vec{c}_k\Big\|\ge 2\delta.
\end{equation}
Thus, by Definition \ref{complete traceability}, $\C$ has $(t,\delta)$-complete traceability if and only if for any two distinct subsets $\C_1,\C_2\subseteq\C$ with $1\le|\C_1|,|\C_2|\le t$, (\ref{distance-1}) holds for any real numbers $\lambda_j,\lambda_k'>0$ such that $\sum_{\vec{c}_j\in\C_1}\lambda_j=\sum_{\vec{c}_k\in\C_2}\lambda_k'=1$, that is, $\mathrm{d}(\mathcal{P}(\C_1),\mathcal{P}(\C_2))\ge 2\delta$ according to (\ref{open convex polytope}). The conclusion follows.
}
\end{IEEEproof}

By Proposition \ref{noisy geometric}, in noisy multimedia fingerprinting, a coalition with size no more than $t$ can be completely traced back if and only if there exists a binary code $\C$ such that for any distinct $\C_1,\C_2\subseteq\C$ with $|\C_1|,|\C_2|\le t$, the condition (\ref{distance}) is satisfied. {\color{black}However, if $\C_1\cap\C_2\neq\emptyset$, we always have $\mathrm{d}(\mathcal{P}(\C_1),\mathcal{P}(\C_2))=0$, and thus (\ref{distance}) does not hold. For example, in Example \ref{3-dim cube}, let $\C_1=\{v_1,v_2,v_6\}$ and $\C_2=\{v_2,v_3\}$. Then $\C_1\cap\C_2\neq\emptyset$ and $\mathcal{P}(\C_1)\cap\mathcal{P}(\C_2)=\emptyset$, but $\mathrm{d}(\mathcal{P}(\C_1),\mathcal{P}(\C_2))=0$ according to (\ref{general distance}).} This immediately implies

%according to (\ref{open convex polytope}) and (\ref{general distance}),

\begin{proposition}
There does not exist any binary code with complete traceability in noisy multimedia fingerprinting.
\end{proposition}

\subsection{Frameproof signature code}

In the previous subsection, we showed from a {\color{black}geometric} viewpoint that there exists no binary code that can trace back to all the colluders in noisy multimedia fingerprinting. In this subsection, we define a binary code with a weaker requirement than that in Definition \ref{complete traceability}, called {\color{black}a} \textit{frameproof signature code}, to provide some security in noisy multimedia fingerprinting. In the literature, Stinson et al. \cite{STW} introduced \textit{secure frameproof codes} in digital fingerprinting to make sure that any illegal copy cannot be generated simultaneously by two disjoint coalition sets. Inspired by this, we introduce the definition of frameproof signature code as follows.

{\color{black}
\begin{definition}\label{secure sig. code}
{\color{black}
Let $\C$ be an $(n,M,2)$ code, $t\ge 2$ be an integer and $\delta>0$ be a real number. $\C$ is an $(n,M,2)$ \textit{$(t,\delta)$-frameproof signature code} if for any two subsets $\C_1,\C_2\subseteq\C$ with $1\le|\C_1|,|\C_2|\le t$ and $\C_1\cap\C_2=\emptyset$, we have
\begin{equation*}
\sum_{\vec{c}_j\in\C_1}\lambda_j\vec{c}_j+\vec{e}\neq\sum_{\vec{c}_k\in\C_2}\lambda_k'\vec{c}_k+\vec{e}'
\end{equation*}
for any $\vec{e},\vec{e}'\in\mathbb{R}^n$ and any real numbers $\lambda_j,\lambda_k'>0$ such that $\|\vec{e}\|,\|\vec{e}'\|<\delta$ and $\sum_{\vec{c}_j\in\C_1}\lambda_j=\sum_{\vec{c}_k\in\C_2}\lambda_k'=1$.
}
\end{definition}
}

According to Definitions~\ref{complete traceability} and~\ref{secure sig. code}, we immediately have a relationship between the frameproof signature code and the code with complete traceability as follows.
\begin{proposition}
Let $\C$ be an $(n,M,2)$ code, $t\ge 2$ be an integer and $\delta>0$ be a real number. If $\C$ has $(t,\delta)$-complete traceability, then $\C$ is a $(t,\delta)$-frameproof signature code, but not vice versa.
\end{proposition}

We provide an equivalent description of frameproof signature codes from a geometric viewpoint.

\begin{proposition}\label{noisy equivalent}
$\C$ is an $(n,M,2)$ $(t,\delta)$-frameproof signature code if and only if for any $\C_1,\C_2\subseteq\C$ with $1\le|\C_1|,|\C_2|\le t$ and $\C_1\cap\C_2=\emptyset$, we have $ \mathrm{d}(\mathcal{P}(\C_1),\mathcal{P}(\C_2))\ge 2\delta$.
\end{proposition}

\begin{example}
It is easy to see from Example \ref{3-dim cube} that $\C=\{(0,0,0),(1,1,0),(1,0,1),(0,1,1)\}$ is a $(2,1/2)$-frameproof signature code.
\end{example}

We remark that although a $(t,\delta)$-frameproof signature code $\C$ cannot identify any coalition set in noisy multimedia fingerprinting, it can guarantee at least two things:
\begin{enumerate}
  \item If $\C_1\subseteq\C$ is a coalition of size at most $t$, then $\C_1$ cannot frame any $\C_2\subseteq\C$ with $|\C_2|\le t$ and $\C_1\cap\C_2=\emptyset$ since they cannot create the same forged copy $\tilde{\vec{y}}$.
  \item If $\C_1\subseteq\C$ is a coalition of size at most $t$ and $\tilde{\vec{r}}$ is the corresponding output of noisy multimedia fingerprinting channel, then any $\C_2\subseteq\C$ with $|\C_2|\le t$ and $\mathrm{d}(\{\tilde{\vec{r}}\},\mathcal{P}(\C_2))<\delta$ contains at least one colluder.
\end{enumerate}

\section{Conclusion}
\label{sec:6}

In this paper, we investigated signature codes for the weighted binary adder channel and collusion-resistant multimedia fingerprinting. We showed the relationships between signature codes and other known combinatorial structures and obtained general and asymptotic upper bounds of $t$-signature codes. We explored the combinatorial properties and derived bounds for $2$-signature codes of constant-weights $2$ and $3$, respectively. Moreover, we provided two explicit constructions for $t$-signature codes which have efficient tracing algorithms. We also introduced two-level signature codes and gave an explicit construction for two-level signature codes. At last, we showed from a geometric viewpoint that there does not exist any binary code with complete traceability for noisy multimedia fingerprinting. As a weaker requirement, to make sure that disjoint coalition sets cannot produce an identical forged copy, we introduced frameproof signature codes for noisy multimedia fingerprinting.

It would be of interest to further improve the bounds for signature codes shown in this paper and find more explicit constructions for signature codes. It would also be of interest to investigate the new type of signature codes introduced in this paper for noisy multimedia fingerprinting.

\section*{Appendix}
%\appendices
\begin{appendices}\label{appendices1}
\begin{IEEEproof}[The proof of $A(4,2)\le 7$]
Assume that $\C\subseteq\{0,1\}^4$ is a $2$-signature code with $|\C|=8$. Let
\begin{equation}\label{proof of $A(4,2)$}
\begin{matrix}
\hspace{0.5cm} \vec{c}_1\hskip 0.8cm \vec{c}_2\hspace{0.5cm} \cdots\hspace{0.5cm} \vec{c}_8\hspace{0.2cm} \\
\C=\begin{bmatrix}
c_1(1) & c_2(1) & \cdots & c_8(1) \\
c_1(2) & c_2(2) & \cdots & c_8(2)  \\
c_1(3) & c_2(3) & \cdots & c_8(3) \\
c_1(4) & c_2(4) & \cdots & c_8(4)
\end{bmatrix}.
\end{matrix}
\end{equation}
By Lemma \ref{sig. code and B code}, it is easy to check that (\ref{proof of $A(4,2)$}) is still a $2$-signature code by exchanging any two rows or any two columns. Moreover, $\{\vec{x}_1, \ldots, \vec{x}_8\}$ is also a $2$-signature code where $\vec{x}_i=(x_i(1),x_i(2),x_i(3),x_i(4))$ and $x_i(j)=1-c_i(j)$ for all $1\le i\le 8$ and $1\le j\le 4$.

First we divide $\C$ into groups with respect to the first and second rows of (\ref{proof of $A(4,2)$}). Let
\begin{align*}
  \C_0=\{\vec{c}_i\in\C: c_i(1)=0\}, & \ \C_1=\{\vec{c}\in\C: c_i(1)=1\}, \\
  \T_0=\{\vec{c}_i\in\C_0:c_i(2)=0\}, & \ \T_1=\{\vec{c}_i\in\C_0: c_i(2)=1\},\\
  \Y_0=\{\vec{c}_i\in\C_1:c_i(2)=0\}, & \ \Y_1=\{\vec{c}\in\C_1:c_i(2)=1\},\\
  \{\vec{a}_1,\vec{a}_2,\vec{a}_3,\vec{a}_4\}=\{0,1\}^2, & \ I=\{2,3,4\}\ \text{and}\ I'=\{3,4\}.
\end{align*}
If $|\C_0|\ge 6$, then $\C_0|_I$ is a $2$-signature code of length $3$ and $|\C_0|_I|=|\C_0|\ge 6$, which contradicts Corollary \ref{$A(2,2)$ and $A(3,2)$} that $A(3,2)=5$. Thus, $|\C_0|\le 5$. Similarly, we also have $|\C_1|\le 5$. Since $\C=\C_0\cup\C_1$ and $|\C|=8$, we only need to discuss the following two cases.
\begin{enumerate}
  \item $|\C_0|=3$ and $|\C_1|=5$. Then $|\Y_0|+|\Y_1|=5$ and $|\T_0|+|\T_1|=3$. By Corollary \ref{$A(2,2)$ and $A(3,2)$} that $A(2,2)=3$, we have $|\Y_0|\le 3$ and $|\Y_1|\le 3$. There are two subcases.
  \begin{enumerate}
  \item[1.1)] $|\Y_0|=2$ and $|\Y_1|=3$. Without loss of generality, we may assume that
  \begin{equation*}
  \C=\begin{bmatrix}
  1 & 1 & 1 & 1 & 1 & 0 & 0 & 0 \\
  0 & 0 & 1 & 1 & 1 & * & * & * \\
  * & * & * & * & * & * & * & * \\
  * & * & * & * & * & * & * & *
  \end{bmatrix}.
  \end{equation*}
  \begin{itemize}
    \item[1.1.a)] If $|\T_0|=3$, then $|\T_0|_{I'}|=3$. Since $|\Y_1|_{I'}|=3$ and $\Y_1|_{I'},\T_0|_{I'}\subseteq\{0,1\}^2$, we have $|\Y_1|_{I'}\cap\T_0|_{I'}|\ge 2$. Without loss of generality, we may assume that
        \begin{equation*}
        \C=\begin{bmatrix}
        1 & 1 & 1 & 1 & 1 & 0 & 0 & 0 \\
        0 & 0 & 1 & 1 & 1 & 0 & 0 & 0 \\
        * & * & * & \vec{a}_1 & \vec{a}_2 & \vec{a}_1 & \vec{a}_2 & *
        \end{bmatrix}.
        \end{equation*}
        Then $\vec{c}_4+\vec{c}_7=\vec{c}_5+\vec{c}_6$, a contradiction to Lemma \ref{sig. code and B code}.
    \item[1.1.b)] If $|\T_0|=2$, we may assume that
        \begin{equation*}
        \C=\begin{bmatrix}
        1 & 1 & 1 & 1 & 1 & 0 & 0 & 0 \\
        0 & 0 & 1 & 1 & 1 & 0 & 0 & 1 \\
        * & * & * & * & * & * & * & * \\
        * & * & * & * & * & * & * & *
        \end{bmatrix}.
        \end{equation*}
        By Lemma \ref{sig. code and B code} and the fact that $|\{0,1\}^2|=4$, we must have $|\T_0|_{I'}\cap\Y_0|_{I'}|\le 1$, $|\T_0|_{I'}\cap\Y_1|_{I'}|=1$ and $|\Y_0|_{I'}\cap\Y_1|_{I'}|=1$. Without loss of generality, we may assume that
        \begin{equation*}
        \C=\begin{bmatrix}
        1 & 1 & 1 & 1 & 1 & 0 & 0 & 0 \\
        0 & 0 & 1 & 1 & 1 & 0 & 0 & 1 \\
        \vec{a}_1 & \vec{a}_2 & \vec{a}_1 & \vec{a}_3 & \vec{a}_4 & \vec{a}_2 & \vec{a}_3 & *
        \end{bmatrix}.
        \end{equation*}
        Then for any choice of $\T_1|_{I'}=\{\vec{a}_j\}$, $j\in\{1,2,3,4\}$, we could have a contradiction to Lemma \ref{sig. code and B code}.
    \item[1.1.c)] If $|\T_0|=1$, we may assume that
        \begin{equation*}
        \C=\begin{bmatrix}
        1 & 1 & 1 & 1 & 1 & 0 & 0 & 0 \\
        0 & 0 & 1 & 1 & 1 & 0 & 1 & 1 \\
        * & * & * & * & * & * & * & * \\
        * & * & * & * & * & * & * & *
        \end{bmatrix}.
        \end{equation*}
        Similarly, we have $|\T_1|_{I'}\cap\Y_0|_{I'}|\le 1$, $|\T_1|_{I'}\cap\Y_1|_{I'}|=1$ and $|\Y_0|_{I'}\cap\Y_1|_{I'}|=1$. Without loss of generality, we may assume that
        \begin{equation*}
        \C=\begin{bmatrix}
        1 & 1 & 1 & 1 & 1 & 0 & 0 & 0 \\
        0 & 0 & 1 & 1 & 1 & 0 & 1 & 1 \\
        \vec{a}_1 & \vec{a}_2 & \vec{a}_1 & \vec{a}_3 & \vec{a}_4 & * & \vec{a}_2 & \vec{a}_3
        \end{bmatrix}.
        \end{equation*}
        If $\T_0|_{I'}=\{\vec{a}_j\}$, $j\in\{1,2,3\}$, then in a similar manner as the case 1.1.b), we could obtain contradictions to Lemma \ref{sig. code and B code}. If $\T_0|_{I'}=\{\vec{a}_4\}$, since $\C$ is a $2$-signature code, we must have
        $$\vec{c}_1+\vec{c}_7\neq\vec{c}_4+\vec{c}_6,\ \vec{c}_2+\vec{c}_8\neq\vec{c}_3+\vec{c}_6,\ \text{and}\ \vec{c}_3+\vec{c}_8\neq\vec{c}_5+\vec{c}_7,$$
        which implies that
        \begin{equation}\label{3 inequalities}
          \vec{a}_1+\vec{a}_2\neq\vec{a}_3+\vec{a}_4,\ \vec{a}_2+\vec{a}_3\neq\vec{a}_1+\vec{a}_4,\ \text{and}\ \vec{a}_1+\vec{a}_3\neq\vec{a}_4+\vec{a}_2.
        \end{equation}
        Since $\{\vec{a}_1,\vec{a}_2,\vec{a}_3,\vec{a}_4\}=\{0,1\}^2$, (\ref{3 inequalities}) cannot be achieved, a contradiction to the assumption.
    \item[1.1.d)] If $|\T_0|=0$, a similar discussion with the case 1.1.a) will lead to a contradiction.
  \end{itemize}
  \item[1.2)] $|\Y_0|=3$ and $|\Y_1|=2$. We can discuss in the same way with the case 1.1) and then also get contradictions.
  \end{enumerate}

  \item $|\C_0|=|\C_1|=4$. Then $|\Y_0|+|\Y_1|=4$ and $|\T_0|+|\T_1|=4$. We only need to consider the case that $|\Y_0|=|\Y_1|=2$ and $|\T_0|=|\T_1|=2$. Without loss of generality, we may assume that
        \begin{equation*}
        \C=\begin{bmatrix}
        1 & 1 & 1 & 1 & 0 & 0 & 0 & 0 \\
        1 & 1 & 0 & 0 & 1 & 1 & 0 & 0 \\
        * & * & * & * & * & * & * & * \\
        * & * & * & * & * & * & * & *
        \end{bmatrix}.
        \end{equation*}
        By Lemma \ref{sig. code and B code}, we have $|\Y_1|_{I'}\cap\Y_0|_{I'}|\le 1$. Then we discuss the following two subcases.
        \begin{enumerate}
          \item[2.1)] $\Y_1|_{I'}\cap\Y_0|_{I'}=\emptyset$. Without loss of generality, we may assume that
              \begin{equation*}
              \C=\begin{bmatrix}
              1 & 1 & 1 & 1 & 0 & 0 & 0 & 0 \\
              1 & 1 & 0 & 0 & 1 & 1 & 0 & 0 \\
              \vec{a}_1 & \vec{a}_2 & \vec{a}_3 & \vec{a}_4 & * & * & * & *
               \end{bmatrix}.
              \end{equation*}
              By Lemma \ref{sig. code and B code}, we must have $|\T_0|_{I'}\cap\T_1|_{I'}|\le 1$ and $|\T_i|_{I'}\cap\Y_j|_{I'}|\le 1$ for $i\neq j\in\{0,1\}$. By discussing the possible choices of $\T_1|_{I'}$, that is, $\{\vec{a}_1,\vec{a}_3\}$, $\{\vec{a}_1,\vec{a}_4\}$, $\{\vec{a}_2,\vec{a}_3\}$ or $\{\vec{a}_2,\vec{a}_4\}$, we could get contradictions to Lemma \ref{sig. code and B code} from $\T_0|_{I'}$.
          \item[2.2)] $|\Y_1|_{I'}\cap\Y_0|_{I'}|=1$. Without loss of generality, we may assume that
              \begin{equation*}
              \C=\begin{bmatrix}
              1 & 1 & 1 & 1 & 0 & 0 & 0 & 0 \\
              1 & 1 & 0 & 0 & 1 & 1 & 0 & 0 \\
              \vec{a}_1 & \vec{a}_2 & \vec{a}_1 & \vec{a}_3 & * & * & * & *
               \end{bmatrix}.
              \end{equation*}
              A similar discussion with the case 2.1) will lead to contradictions.
        \end{enumerate}
\end{enumerate}

Then we have $A(4,2)\le 7$.
\end{IEEEproof}
\end{appendices}

\section*{Acknowledgment}
Miao is grateful to Prof. Itzhak Tamo of Tel Aviv University and Prof. Zhiying Wen of Tsinghua University for their insightful discussions on Section \ref{sec:5}.
All the authors express their sincere thanks to the three anonymous reviewers and Prof. Joerg Kliewer, the Associate Editor, for their valuable comments and suggestions which greatly improved this paper.

\ifCLASSOPTIONcaptionsoff
  \newpage
\fi

\end{document}